\documentclass[aps,prc,preprint,floatfix]{revtex4}

\usepackage{amsmath}
\usepackage{graphicx}
\DeclareGraphicsExtensions{.ps}

\def\gtorder{\mathrel{\raise.3ex\hbox{$>$}\mkern-14mu
	\lower0.6ex\hbox{$\sim$}}}
\def\ltorder{\mathrel{\raise.3ex\hbox{$<$}\mkern-14mu
	\lower0.6ex\hbox{$\sim$}}}
\newcommand{\sW}{\sin^2\theta_W}

\begin{document}

\title{Finite-$Q^2$ Corrections to Parity-Violating DIS}

\author{T. Hobbs}
\affiliation{The University of Chicago, 5801 South Ellis Avenue,
        Chicago, IL 60637}
\author{W. Melnitchouk}
\affiliation{Jefferson Lab, 12000 Jefferson Avenue,
        Newport News, VA 23606}

% \date{\today}

\begin{abstract}
Parity-violating deep inelastic scattering (PVDIS) has been proposed
as an important new tool to extract the flavor and isospin dependence
of parton distributions in the nucleon.
We discuss finite-$Q^2$ effects in PVDIS asymmetries arising from
subleading kinematical corrections and longitudinal contributions to
the $\gamma Z$ interference.
For the proton, these need to be accounted for in order to accurately
extract the $d/u$ ratio at large $x$; for the deuteron they are important
to consider when searching for evidence of charge symmetry violation in
parton distributions or signals for physics beyond the standard model.
We further explore the dependence of PVDIS asymmetries for polarized
targets on the $u$ and $d$ helicity distributions at large $x$.
\end{abstract}

\maketitle

%%%%%%%%%%%%%%%%%%%%%%%%%%%%%%%%%%%%%%%%%%%%%%%%%%%%%%%%%%%%%%%%%%%%%%%%%
\section{Introduction}
\label{sec:intro}

The scattering of highly energetic leptons from nucleon targets has
over the years provided a wealth of information on the nucleon quark
and gluon (or parton) substructure.
Most of the information has come from electromagnetic deep inelastic
scattering (DIS) of electrons (or muons), while neutrino DIS has
yielded complementary constraints on valence and sea parton
distribution functions (PDFs) via the weak current.

A relatively unexplored method of measuring PDFs is through the
interference of electromagnetic and weak currents, which in
principle selects a unique combination of quark flavors.
This involves measuring the small $\gamma$--$Z^0$ interference
amplitude in the neutral current DIS of a polarized electron
from a hadron $h$, $\vec{e}\ h \to e\ X$.
Because the axial current is sensitive to the polarization of the
incident electron, measurement of the asymmetry between left- and
right-hand polarized electrons is proportional to the $\gamma$--$Z^0$
interference amplitude.

In fact, in the 1970s parity-violating deep inelastic scattering (PVDIS)
on the deuteron provided an important early confirmation of the standard
model of particle physics \cite{Prescott,Cahn}.
Three decades on, experimental techniques are sophisticated enough
now to enable left-right asymmetries to be measured to a few 
parts-per-million, and current facilities will be able to improve
the statistics of the earlier experiments by an order of magnitude
\cite{JLab6,JLab12}.

For parity-violating scattering from an isoscalar deuteron, the
dependence of the left-right asymmetry on PDFs cancels in the
parton model, so that the asymmetry is determined entirely by
the Weinberg angle, $\theta_W$.
In the SU(2)$\times$U(1) electroweak theory, the Lagrangian
corresponding to the parity-violating (PV) lepton-quark interaction
(for two quark flavors) is given by \cite{PDG,Bj}:
\begin{equation}
{\cal L}^{\rm PV}
= \frac{G_F}{\sqrt{2}}
  \left[
    \bar{e} \gamma^\mu \gamma_5e
    \left( C_{1u} \bar{u} \gamma_\mu u
	 + C_{1d} \bar{d} \gamma_\mu d
    \right)
  + \bar{e} \gamma^\mu e
    \left( C_{2u} \bar{u} \gamma_\mu \gamma_5 u
	 + C_{2d} \bar{d} \gamma_\mu \gamma_5 d
    \right)
 \right]\ ,
\end{equation}
where $G_F$ is the Fermi coupling constant, and the electroweak
couplings at tree level are:
\begin{subequations}
\label{eq:C12}
\begin{align}
C_{1u}
&=\ g^e_A \cdot g^u_V\ 
 =\ -\frac{1}{2}\ +\ \frac{4}{3}\sW\ , \\
C_{1d}
&=\ g^e_A \cdot g^d_V\
 =\ \ \ \, \frac{1}{2}\ -\ \frac{2}{3}\sW\ ,\\
C_{2u}
&=\ g^e_V \cdot g^u_A\
 =\ -\frac{1}{2}\ +\ 2\sW\ ,	\\
C_{2d}
&=\ g^e_V \cdot g^d_A\
 =\ \ \ \, \frac{1}{2}\ -\ 2\sW\ .
\end{align}
\end{subequations}
With our conventions the vector and axial vector couplings of the
charged lepton are $g^e_V = -1 + 4\sW$ and $g^e_A = +1$, respectively.
The vector couplings of the quarks are given by
$g^u_V = -1/2 + (4/3)\sW$ and
$g^d_V =  1/2 - (2/3)\sW$,
while the quark axial vector couplings are
$g^u_A =  1/2$ and
$g^d_A = -1/2$, respectively.
The deuteron asymmetry is therefore a sensitive test of effects beyond
the parton model, such as higher twist contributions, or of more exotic
effects such as charge symmetry violation in PDFs or new physics beyond
the standard model.

More recently it has been suggested that PVDIS can be used to probe
parton distribution functions in the largely unmeasured region of
high Bjorken-$x$ \cite{Souder,SLAC}.
In particular, the PVDIS asymmetry for a proton is proportional
to the ratio of $d$ to $u$ quark distributions at large $x$.
Current determinations of the $d/u$ ratio rely heavily on inclusive
proton and deuteron DIS data, and there are large uncertainties in
the nuclear corrections in the deuteron at high $x$ \cite{NP}.
While novel new methods have been suggested to minimize the nuclear
uncertainties \cite{A3,BONUS,Comment}, the of use a proton target alone
would avoid the problem altogether.

In this paper we critically examine the accuracy of the parton model
predictions for the PVDIS asymmetries in realistic experimental
kinematics at finite $Q^2$.
In particular, in Sec.~\ref{sec:PVDIS} we provide a complete set of
formulas for cross sections and asymmetries for scattering polarized
leptons from unpolarized targets, including finite-$Q^2$ effects.
PVDIS from the proton is discussed in Sec.~\ref{sec:p}, where we test
the sensitivity of the extraction of the $d/u$ ratio at large $x$ to
finite-$Q^2$ corrections.
One of the main uncertainties in the calculation is the ratio of
longitudinal to transverse cross sections for the $\gamma$--$Z^0$
interference, for which no empirical information currently exists.
We provide some numerical estimates of the possible dependence of the
left-right asymmetry on this ratio.

For deuteron targets, we examine in Sec.~\ref{sec:d} how the asymmetry
is modified in the presence of finite-$Q^2$ corrections, and where
these can pose significant backgrounds for extracting standard model
signals.
Finally, we explore in Sec.~\ref{sec:pol} the possibility of using
PVDIS with polarized targets to constrain quark helicity distributions
at large $x$.
A comprehensive discussion of polarized PVDIS in the parton model was
previously given by Anselmino {\em et al.} \cite{Ans}; here we perform
a numerical survey of the sensitivity of polarized PVDIS asymmetries
to spin-dependent PDFs.
In Sec.~\ref{sec:conc} we make concluding remarks and outline future
work.

%%%%%%%%%%%%%%%%%%%%%%%%%%%%%%%%%%%%%%%%%%%%%%%%%%%%%%%%%%%%%%%%%%%%%%%%%
\section{Parity-Violating Deep Inelastic Scattering}
\label{sec:PVDIS}

In this section we outline the formalism relevant for parity-violating
deep inelastic scattering of an electron (four-momentum $l$) from a
nucleon target ($p$) to a scattered electron ($l'$) and hadronic
debris ($p_X$), via the exchange of a virtual photon or $Z^0$-boson
($q$).
We discuss the general decomposition of the hadronic tensor, and
provide formulas for the PV asymmetry in terms of structure functions,
and in the parton model in terms of PDFs.

% .......................................................................
\subsection{Hadronic Tensor}

We begin with the differential cross section for inclusive
electron--nucleon scattering, which in general can be written as
the squared sum of the $\gamma$- and $Z^0$-exchange amplitudes.
We will consider contributions to the cross section from the pure
$\gamma$ exchange amplitude and the $\gamma$--$Z$ interference;
the purely weak $Z^0$ exchange contribution to the cross section
is strongly suppressed relative to these and can be neglected.

Formally, the cross section can be written in terms of products
of leptonic and hadronic tensors as \cite{Ans,TWbook}:
\begin{equation}
\label{eq:dsig}
\frac{d^2\sigma}{d\Omega dE'}
= {\alpha^2 \over Q^4}
  {E' \over E}
  \left( L^\gamma_{\mu \nu} W^{\mu \nu}_\gamma\
     +\ {G_F \over 4 \sqrt{2} \pi \alpha}
         L^{\gamma Z}_{\mu \nu} W^{\mu\nu}_{\gamma Z}
  \right)\ ,
\end{equation}
where $E$ and $E'$ are the (rest frame) electron energies,
$Q^2$ is (minus) the four-momentum transfer squared, and
$\alpha$ is the electromagnetic fine structure constant.
The lepton tensor for the interference current in Eq.~(\ref{eq:dsig})
is given by:
\begin{equation}
\label{eq:LgZ}
L^{\gamma Z}_{\mu\nu} = (g^e_V + \lambda g^e_A)\ L^\gamma_{\mu\nu}\ ,
\end{equation}
with $\lambda = +1 (-1)$ for positive (negative) initial lepton
helicity, and the purely electromagnetic tensor is given by:
\begin{equation}
L^\gamma_{\mu\nu}
= 2 \left( l_\mu l'_\nu + l'_\mu l_\nu - l \cdot l'g_{\mu \nu}
	 + i \lambda\varepsilon_{\mu\nu\alpha\beta}\ l^\alpha l'^\beta
    \right)\ .
\end{equation}

The hadronic tensors for the electromagnetic and interference
contributions are given by:
\begin{eqnarray}
W_{\mu\nu}^{\gamma(\gamma Z)}
&=& {1 \over 2M}
  \sum_X 
  \left\{
    \langle X | J_\mu^{\gamma(Z)} | N \rangle^*
    \langle X | J_\nu^{\gamma}    | N \rangle
  + \langle X | J_\mu^{\gamma}    | N \rangle^*
    \langle X | J_\nu^{\gamma(Z)} | N \rangle
  \right\}		\nonumber\\
& & \times (2\pi)^3\ \delta(p_X - p - q)\ ,
\end{eqnarray}
where $M$ is the nucleon mass, and $J_\mu^{\gamma(Z)}$ is the
electromagnetic (weak) hadronic current.
In general, the hadronic tensor for a nucleon with spin four-vector
$S^\mu$ can be written in terms of 3 spin-independent and 5
spin-dependent structure functions \cite{Ans}:
\begin{eqnarray}
W_{\mu\nu}^i
&=& - \frac{g_{\mu\nu}}{M}\ F_1^i\
 +\ \frac{p_\mu p_\nu}{M\ p\cdot q}\ F_2^i\
 +\ \frac{i \varepsilon_{\mu\nu\alpha\beta} p^\alpha q^\beta}
	 {2M p\cdot q}\ F_3^i				\nonumber\\
&+& \frac{i \varepsilon_{\mu\nu\alpha\beta}}{p\cdot q}
    \left( q^\alpha S^\beta\ g_1^i
	 + 2x p^\alpha S^\beta\ g_2^i
    \right)\
 -\ \frac{p_\mu S_\nu + S_\mu p_\nu}{2 p\cdot q}\ g_3^i		\\
&+& \frac{S \cdot q\ p_\mu p_\nu}{(p \cdot q)^2}\ g_4^i
 +\ \frac{S \cdot q\ g_{\mu\nu}}{p \cdot q}\ g_5^i\ ,
\nonumber
\end{eqnarray}
for both the electromagnetic ($i=\gamma$) and interference
($i=\gamma Z$) currents.
Each of the structure functions generally depend on two variables,
usually taken to be $Q^2$ and the Bjorken scaling variable
$x = Q^2/2M\nu$, where $\nu$ is the energy transfer.

Below we will consider scattering of a polarized electron from an
unpolarized hadron target, in which only the spin-independent
structure functions $F_{1-3}^{\gamma Z}$ enter.
Asymmetries resulting from scattering of an unpolarized electron
beam from a polarized target, which are sensitive to the
spin-dependent structure functions $g_{1-5}^{\gamma Z}$, will be
discussed in Sec.~\ref{sec:pol}.

% .......................................................................
\subsection{Beam Asymmetries}

The PV interference structure functions can be isolated by constructing
an asymmetry between cross sections for right- ($\sigma_R$) and left-hand
($\sigma_L$) polarized electrons:
\begin{equation}
\label{eq:APVdef}
A^{\rm PV} = \frac{\sigma_R - \sigma_L}{\sigma_R + \sigma_L}\ ,
\end{equation}
where $\sigma \equiv d^2\sigma/d\Omega dE'$.
Since the purely electromagnetic and purely weak cross sections
do not contribute to the asymmetry for $Q^2 \ll M_Z^2$, the numerator
is sensitive only to the $\gamma$--$Z$ interference term.
The denominator, on the other hand, is dominated by the purely 
electromagnetic component.
In terms of structure functions, the PVDIS asymmetry can be written:
\begin{equation}
A^{\rm PV}
= - \left( {G_F Q^2 \over 4 \sqrt{2} \pi \alpha} \right)
  { g^e_A
    \left( 2xy F_1^{\gamma Z} - 2 [1 - 1/y + xM/E] F_2^{\gamma Z} \right)
  + g^e_V\
    x (2-y) F_3^{\gamma Z}
  \over
    2xy F_1^\gamma - 2 [1 - 1/y + xM/E] F_2^\gamma
  }\ .
\end{equation}
where $y=\nu/E$ is the lepton fractional energy loss.

In the Bjorken limit ($Q^2, \nu \to \infty$, $x$ fixed),
the interference structure functions $F_1^{\gamma Z}$ and
$F_2^{\gamma Z}$ are related by the Callan-Gross relation,
$F_2^{\gamma Z} = 2x F_1^{\gamma Z}$, similar to the electromagnetic
$F_{1,2}^\gamma$ structure functions \cite{Ans}.
At finite $Q^2$, however, corrections to this relation are usually
parameterized in terms of the ratio of the longitudinal to transverse
virtual photon cross sections:
\begin{equation}
R^{\gamma (\gamma Z)}\
\equiv\ \frac{\sigma_L^{\gamma (\gamma Z)}}{\sigma_T^{\gamma (\gamma Z)}}\ 
=\ r^2 \frac{F_2^{\gamma (\gamma Z)}}{2x F_1^{\gamma (\gamma Z)}} - 1\ ,
\end{equation}
for both the electromagnetic ($\gamma$) and interference ($\gamma Z$)
contributions, with
\begin{equation}
r^2 = 1 + {Q^2 \over \nu^2} = 1 + {4 M^2 x^2 \over Q^2}\ .
\end{equation}

In terms of this ratio, the PVDIS asymmetry can be written more
simply as:
\begin{equation}
A^{\rm PV}
= - \left( {G_F Q^2 \over 4 \sqrt{2} \pi \alpha} \right)
  \left[ g^e_A\ Y_1\ \frac{F_1^{\gamma Z}}{F_1^{\gamma}}\
     +\ {g^e_V \over 2}\ Y_3\ \frac{F_3^{\gamma Z}}{F_1^{\gamma}} 
  \right]\ ,
\label{eq:APV}
\end{equation}
where the functions $Y_{1,3}$ parameterize the dependence on $y$
and on the $R$ ratios:
\begin{subequations}
\label{eq:Y}
\begin{align}
\label{eq:Y1}
Y_1
% &= \frac{ 1+(1-y)^2-y^2 (1-r^2/(1+R^{\gamma Z})) - 2xyM/E }
%	{ 1+(1-y)^2-y^2 (1-r^2/(1+R^\gamma)) - 2xyM/E } 
%%%
%%% 2xyM/E => xyM/E CORRECTED
%%% [see Eqs. (21) of Brady et al., PRD 84, 074008 (2011)]
%%%
%
&= \frac{ 1+(1-y)^2-y^2 (1-r^2/(1+R^{\gamma Z})) - xyM/E }
	{ 1+(1-y)^2-y^2 (1-r^2/(1+R^\gamma)) - xyM/E } 
   \left( \frac{1+R^{\gamma Z}}{1+R^\gamma} \right)\ ,	\\
\label{eq:Y3}
Y_3
&= \frac{ 1-(1-y)^2 }
%	{ 1+(1-y)^2-y^2 (1-r^2/(1+R^\gamma)) - 2xyM/E }
	{ 1+(1-y)^2-y^2 (1-r^2/(1+R^\gamma)) - xyM/E }
   \left( \frac{r^2}{1+R^\gamma} \right)\ .
\end{align}
\end{subequations}
In the Bjorken limit, the kinematical ratio $r^2 \to 1$,
while the longitudinal cross section vanishes relative to the
transverse, $R^i \to 0$, for both $i = \gamma$ and $\gamma Z$.
For kinematics relevant to future experiments ($Q^2 \sim$~few GeV$^2$,
$\nu \sim$~few GeV), the factor $xyM/E$ provides a small correction,
and can for practical purposes be dropped.
In this case the functions $Y_1$ and $Y_3$ have the familiar limits
\cite{Prescott}:
\begin{subequations}
\label{eq:Ybj}
\begin{align}
\label{eq:Y1bj}
Y_1 &\to 1\ ,	\\
\label{eq:Y3bj}
Y_3 &\to \frac{1-(1-y)^2}{1+(1-y)^2}\ \equiv\ f(y)\ .
\end{align}   
\end{subequations}
Typically the contribution from the $Y_3$ term is much smaller
than from the $Y_1$ term because $g_V^e \ll g_A^e$, although for
quantitative comparisons it needs to be included.

% .......................................................................
\subsection{Electroweak Structure Functions}
\label{ssec:SFs}

The PVDIS asymmetry $A^{\rm PV}$ can be evaluated from knowledge
of the electromagnetic and interference structure functions.
At leading twist of the electroweak structure functions can be
expressed in terms of PDFs.
For reference these are listed (at leading order in $\alpha_s$)
as follows:
\begin{subequations}
\label{eq:Fg}
\begin{align}
F_1^\gamma(x) &= \frac{1}{2} \sum_q e_q^2\ (q(x) + \bar{q}(x))\ , \\
F_2^\gamma(x) &= 2x F_1^\gamma(x)\ ,
\end{align}
\end{subequations}
for the pure electromagnetic case, while
\begin{subequations}
\label{eq:FgZ}
\begin{align}
F_1^{\gamma Z}(x) &= \sum_q e_q\ g_V^q\ (q(x) + \bar{q}(x))\ , 	\\
F_2^{\gamma Z}(x) &= 2x F_1^{\gamma Z}(x)\ ,			\\
F_3^{\gamma Z}(x) &= 2 \sum_q e_q\ g_A^q\ (q(x) - \bar{q}(x))\ ,
\end{align}
\end{subequations}
are the structure functions for the weak-electromagnetic interference,
where the quark $q$ and antiquark $\bar q$ distributions are defined
with respect to the proton.

In terms of PDFs, the PV asymmetry in Eq.~(\ref{eq:APV}) can be
written as:
\begin{equation}
A^{\rm PV}
= - \left( {G_F Q^2 \over 4 \sqrt{2} \pi \alpha} \right)
    \left( Y_1\ a_1\ +\ Y_3\ a_3 \right)\ ,
\end{equation}
where the vector term $a_1$ is given by:
\begin{subequations}
\label{eq:a13}
\begin{equation}
\label{eq:a1}
a_1 = \frac{2 \sum_q  e_q\ C_{1q}\ (q+\bar q)}
	   {  \sum_q  e_q^2\       (q+\bar q)}\ ,
\end{equation}
while the axial vector term is:
\begin{equation}
\label{eq:a3}
a_3 = \frac{2 \sum_q  e_q\ C_{2q}\ (q-\bar q)}
	   {  \sum_q  e_q^2\       (q+\bar q)}\ .
\end{equation}
\end{subequations}
In this analysis we will focus on the large-$x$ region dominated by
valence quarks, so that the effects of sea quark will be negligible.

At finite $Q^2$, corrections to the parton model expressions appear in
the form of perturbatively generated $\alpha_s$ corrections, target mass
corrections \cite{TMC}, as well as higher twist ($1/Q^2$ suppressed)
effects.
Some higher twist effects in PVDIS have previously been investigated in
the literature \cite{HT}.
One should also note that at large $x$ perturbative calculations beyond
leading order can become unstable and threshold resummations need to be 
performed \cite{resum}.

A detailed study of each of these corrections will be published
elsewhere \cite{future}; in the present study we focus on the
finite-$Q^2$ effects on the asymmetry arising from non-zero values
of $R^{\gamma(\gamma Z)}$, which to date have not been considered.
While data and phenomenological parametrizations are available for
$R^\gamma$ \cite{R1990,R1998,RJLab}, currently no empirical
information exists on $R^{\gamma Z}$.
In our numerical estimates below, we shall consider a range of
possible behaviors for $R^{\gamma Z}$ and examine its effect on
$A^{\rm PV}$.

%%%%%%%%%%%%%%%%%%%%%%%%%%%%%%%%%%%%%%%%%%%%%%%%%%%%%%%%%%%%%%%%%%%%%%%%%
\section{PVDIS on the Proton}
\label{sec:p}

Parity-violating DIS on a proton target has recently been discussed
as a means of constraining the ratio of $d$ to $u$ quark distributions
at large $x$ \cite{Souder}.
At present the $d/u$ ratio is essentially unknown beyond $x \sim 0.6$
due to large uncertainties in the nuclear corrections in the deuteron,
which is the main source of information on the $d$ quark distribution
\cite{NP,Comment}.
Several new approaches to determining $d/u$ at large $x$ have been
proposed, for example using spectator proton tagging in semi-inclusive
DIS from the deuteron \cite{BONUS}, or through a ratio of $^3$He and
$^3$H targets to cancel the nuclear corrections \cite{A3}.
The virtue of the PVDIS method is that, rather than using different
hadrons (or nuclei) to select different flavors, here one uses (the
interference of) different gauge bosons to act as the flavor filter,
thereby avoiding nuclear uncertainties altogether.

\begin{figure}[t]
\includegraphics[height=8cm]{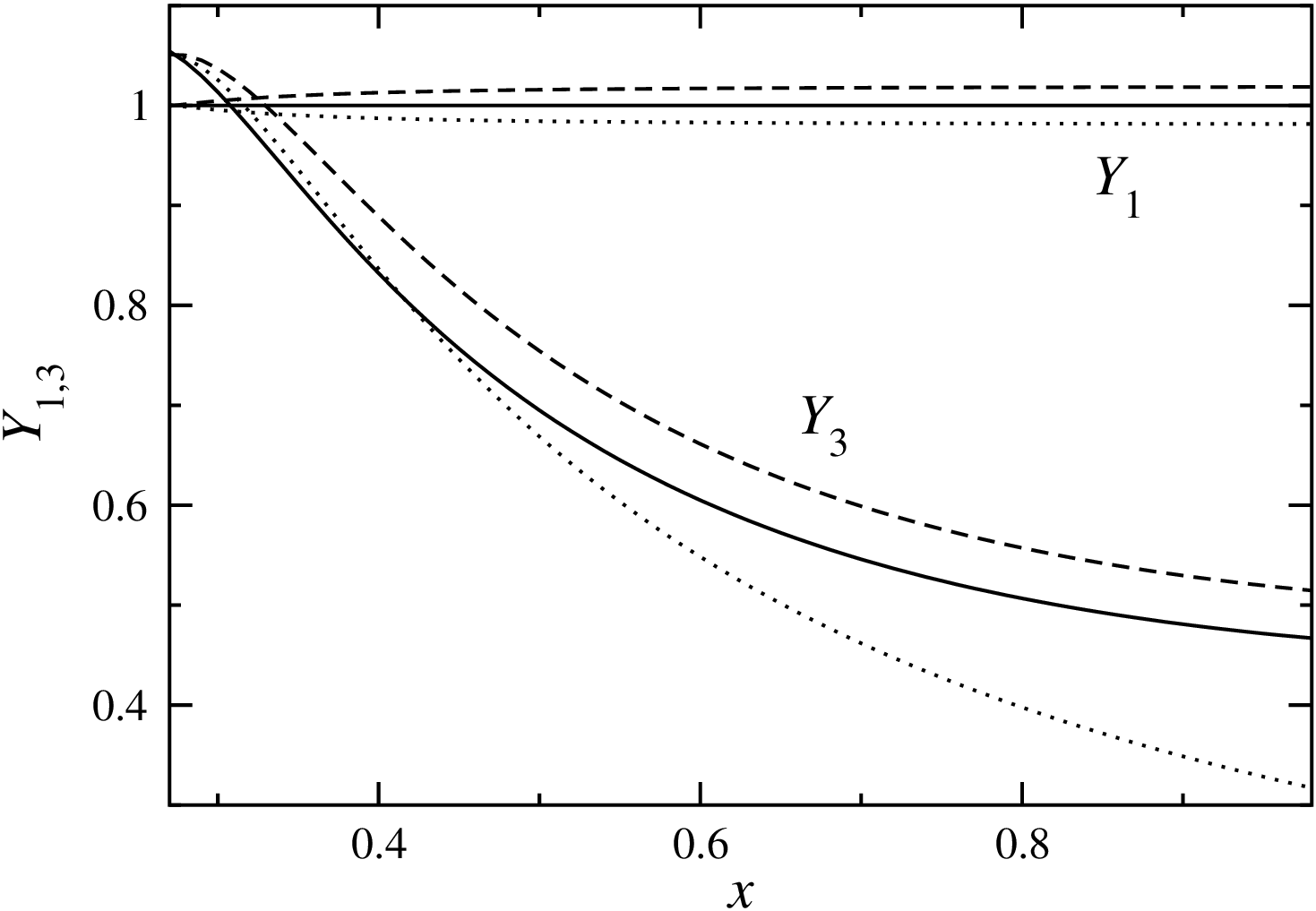}
\caption{$Y_1$ and $Y_3$ as a function of $x$, for $Q^2=5$~GeV$^2$
	and $E=10$~GeV.
	For $Y_1$, the solid line (at $Y_1=1$) corresponds to
	$R^{\gamma Z} = R^\gamma$ \cite{R1990}, while the dashed 
	(dotted) curves around it represent $+(-) 20\%$ deviations
	of $R^{\gamma Z}$ from $R^\gamma$.
	For $Y_3$, the Bjorken limit result ($R^\gamma=0, r^2=1$)
	is given by the dotted curve, the dashed has $R^\gamma=0$
	but $r^2 \neq 1$, while the solid represents the full result.}
\label{fig:Y13}
\end{figure}

In the valence region at large $x$, the PV asymmetry is sensitive
to the valence $u$ and $d$ quark distributions in the proton.
Here the functions $a_1$ and $a_3$ in Eqs.~(\ref{eq:a13}) for the
proton can be simplified to:
\begin{subequations}
\begin{equation}
\label{eq:a1p}
a_1^p = \frac{12 C_{1u} - 6 C_{1d}\ d/u}{4 + d/u}\ ,
\end{equation}
and
\begin{equation}
\label{eq:a3p}
a_3^p = \frac{12 C_{2u} - 6 C_{2d}\ d/u}{4 + d/u}\ .
\end{equation}  
\end{subequations}
This reveals that both $a_1^p$ and $a_3^p$ depend on the $d/u$ quark
distribution ratio.

\begin{figure}[t]
\includegraphics[height=5.5cm]{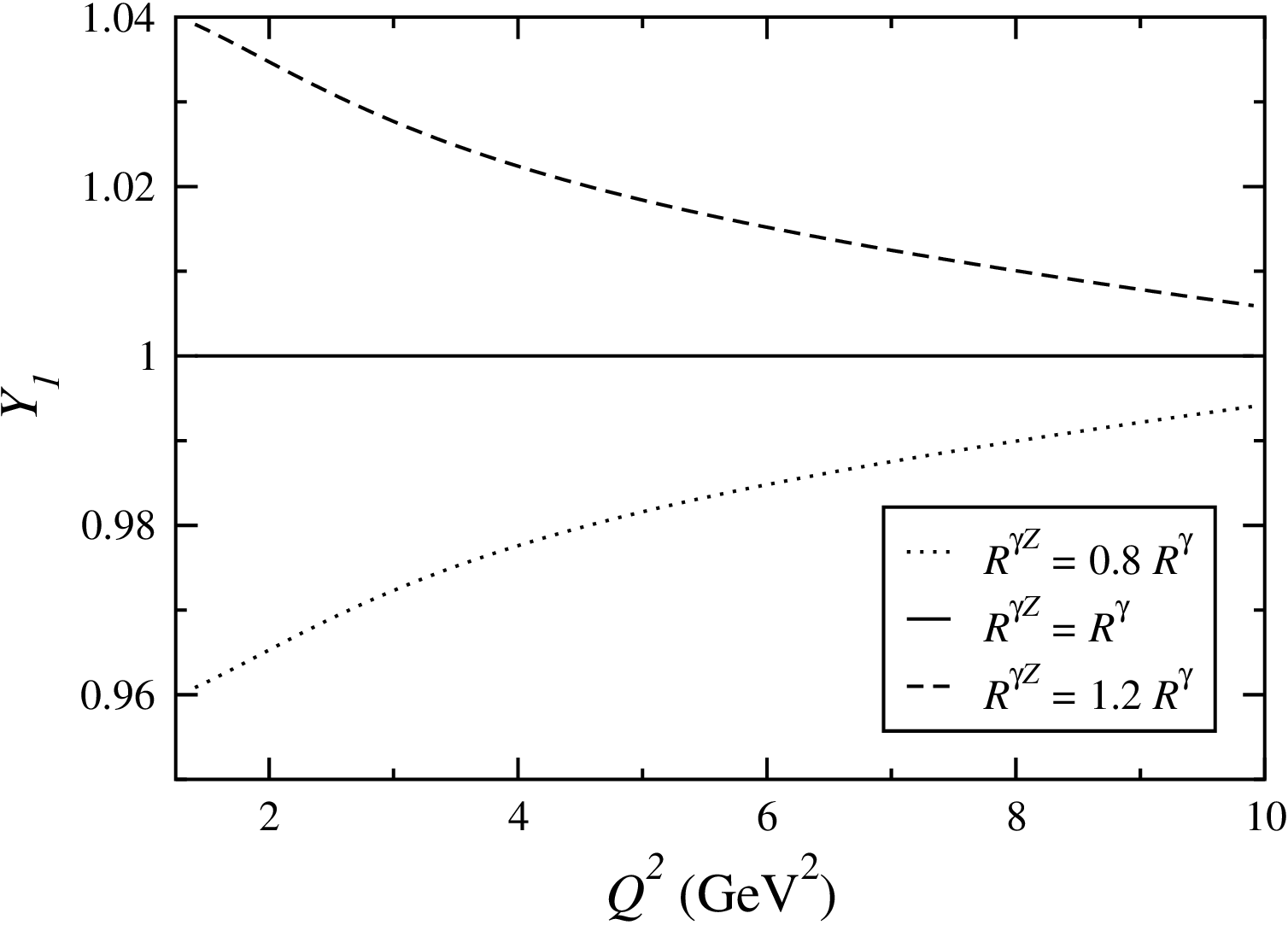}
\includegraphics[height=5.5cm]{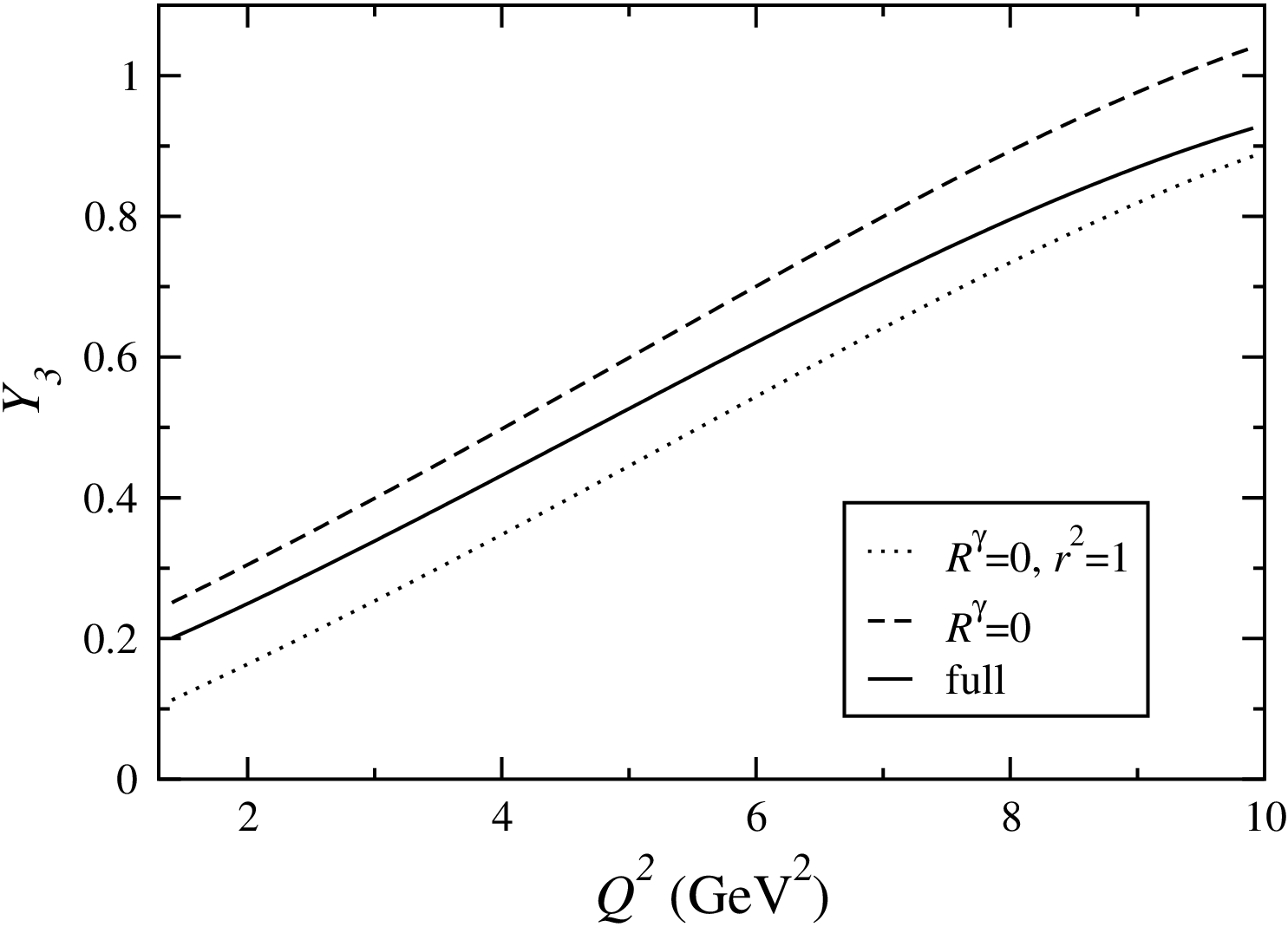}
\caption{$Y_1$ and $Y_3$ as a function of $Q^2$, for $x=0.7$
	and $E=10$~GeV:
	(a) Dependence of $Y_1$ on $R^{\gamma Z}$, for
	$R^{\gamma Z} = 0.8 R^\gamma$ (dotted),
	$R^{\gamma Z} = R^\gamma$ (solid), and
	$R^{\gamma Z} = 1.2 R^\gamma$ (dashed).
	(b) Dependence of $Y_3$ on $R^\gamma$,
	in the Bjorken limit ($R^\gamma=0, r^2=1$) (dotted),
	with $R^\gamma = 0$ but $r^2 \neq 1$ (dashed),
	and full result (solid).}
\label{fig:YQ}
\end{figure}

To explore the relative sensitivity of the proton asymmetry
$A^{\rm PV}_p$ to the vector and axial vector terms, in
Fig.~\ref{fig:Y13} we show the functions $Y_1$ and $Y_3$ for
the proton as a function of $x$, evaluated at $Q^2 = 5$~GeV$^2$,
for a beam energy $E = 10$~GeV (which we will assume throughout).
For $Y_1$, the solid line (at $Y_1=1$) corresponds to
$R^{\gamma Z} = R^\gamma$, while the dashed (dotted) curves around it
represent $+(-) 20\%$ deviations of $R^{\gamma Z}$ from $R^\gamma$
(see below).
For $Y_3$, the Bjorken limit result ($R^\gamma=0, r^2=1$) is given by
the dotted curve, the dashed curve has $R^\gamma=0$ but $r^2 \neq 1$,
while the solid represents the full result with $R^\gamma \neq 0$
and $r^2 \neq 1$.
In all cases we use $R^\gamma$ from the parametrization of
Ref.~\cite{R1990}.
The results with the parametrization of Ref.~\cite{R1998} are very
similar, and are consistent within the quoted uncertainties.

Note that at fixed $Q^2$, the large-$x$ region also corresponds to
low hadronic final state masses $W$, so that with increasing $x$ one
eventually encounters the resonance region at $W \lesssim 2$~GeV.
For $Q^2 = 5$~GeV$^2$ this occurs at $x \approx 0.62$, and for
$Q^2 = 10$~GeV$^2$ at $x \approx 0.76$.
This may introduce an additional source of uncertainty in the
extraction of the PV asymmetry at large $x$, arising from possible
higher twist corrections to structure functions.
In actual experimental conditions, the value of $Q^2$ can be varied
with $x$ to ensure that the resonance region is excluded from the
data analysis.
For the purposes of illustrating the finite-$Q^2$ effects in our
analysis, we shall fix $Q^2$ at the low end of values attainable
with an energy of $E = 10$~GeV, namely $Q^2 = 5$~GeV$^2$.

The relative roles played by the vector and axial vector terms at
different $Q^2$ values is illustrated in Fig.~\ref{fig:YQ}, where
$Y_1$ and $Y_3$ are plotted as a function of $Q^2$ at a fixed $x = 0.7$.
In Fig.~\ref{fig:YQ}(a) we show the dependence of $Y_1$ on the
interference ratio $R^{\gamma Z}$.
With $R^{\gamma Z} = R^\gamma$ the result is unity, as expected
from Eq.~(\ref{eq:Y1}).
Varying $R^{\gamma Z}$ by $\pm 20\%$ relative to $R^\gamma$
\cite{R1990} results in an $\approx 4\%$ shift at
$Q^2 \sim 1$~GeV$^2$, decreasing to $< 1\%$ for
$Q^2 \sim 10$~GeV$^2$.

For the axial vector contribution, in Fig.~\ref{fig:YQ}(b) we show
$Y_3$ under various kinematical approximations, namely in the Bjorken
limit ($R^\gamma=0, r^2=1$), for $R^\gamma = 0$ but $r^2 \neq 1$,
and the full result.
The differences between the full and Bjorken limit results are of the
order 40\% at $Q^2 = 5$~GeV$^2$ and $\sim 20\%$ at $Q^2 = 10$~GeV$^2$.
The rise in $Y_3$ with $Q^2$ is kinematical, since $y \sim \nu \sim Q^2$
for fixed $x$ and $E$.
Because the axial contribution is suppressed relative to the vector term
in $A^{\rm PV}$, $g_V^e \ll g_A^e$, the uncertainty in $A^{\rm PV}$
arising from $Y_3$ will be less significant.
Numerically, the ratio $a_3^p/a_1^p$ of the axial to vector terms,
using the CTEQ6 \cite{CTEQ} parametrization of the PDFs, ranges from
$0.21 - 0.24$ for $0.4 < x < 0.9$.
Although the axial vector $a_3^p$ term is small, it is nevertheless
important to take into account in precision determinations of
$A_p^{\rm PV}$.

\begin{figure}[t]
\includegraphics[height=9cm]{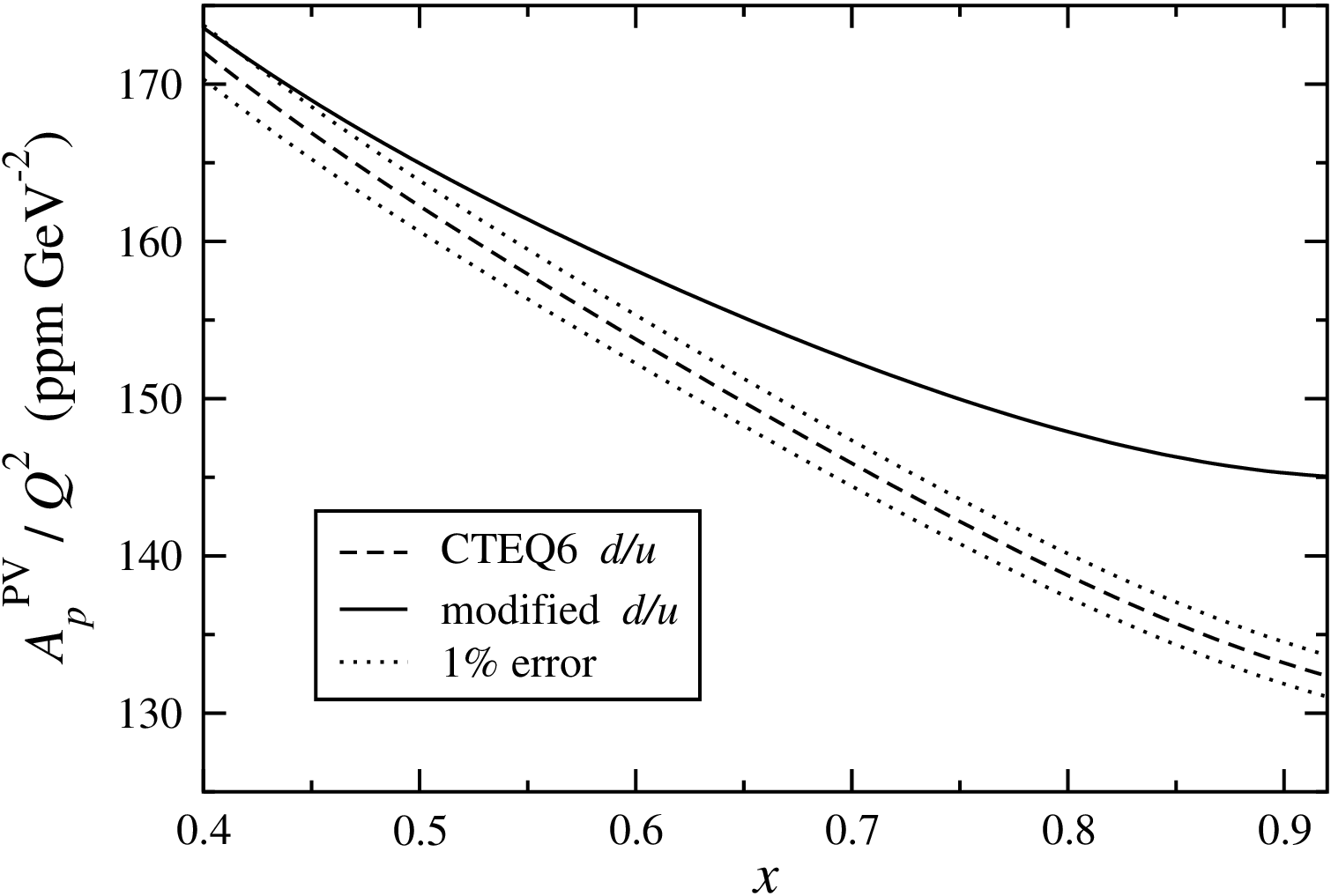}
\caption{Proton PV asymmetry $A^{\rm PV}_p/Q^2$ as a function of $x$,
	for $Q^2 = 5$~GeV$^2$, in parts per million (ppm)
	$\cdot$~GeV$^{-2}$.
	The prediction with the standard CTEQ6 PDFs (dashed) is
	compared with that using a modified $d/u$ ratio at large $x$
	(solid).
	A $\pm 1\%$ uncertainty band (dotted) is shown around the
	standard CTEQ6 prediction.}
\label{fig:APV_du}
\end{figure}

The sensitivity of the proton asymmetry $A^{\rm PV}_p$, measured in
parts per million (ppm), to the $d/u$ ratio is illustrated in
Fig.~\ref{fig:APV_du} as a function of $x$, for $Q^2 = 5$~GeV$^2$,
where $A^{\rm PV}_p/Q^2$ is shown.
Here we assume that $R^{\gamma Z} = R^\gamma$, so that the coefficient
$Y_1$ in the vector term is unity.
For the $u$ and $d$ distributions we use the CTEQ6 PDF set \cite{CTEQ},
in which the $d/u$ ratio vanishes as $x \to 1$, along with a modified
$d/u$ ratio which has a finite $x \to 1$ limit of 0.2 \cite{NP},
$d/u \to d/u + 0.2\ x^2\ \exp(-(1-x)^2)$ \cite{Peng}, motivated by
theoretical counting rule arguments \cite{FJ}.
Also shown (dotted band around the CTEQ6 prediction) is a $\pm 1\%$
uncertainty, which is a conservative estimate of what may be expected
experimentally at JLab with 12~GeV \cite{Souder,JLab12}.
The results indicate that a signal for a larger $d/u$ ratio would
be clearly visible above the experimental errors.

\begin{figure}[t]
\includegraphics[height=9cm]{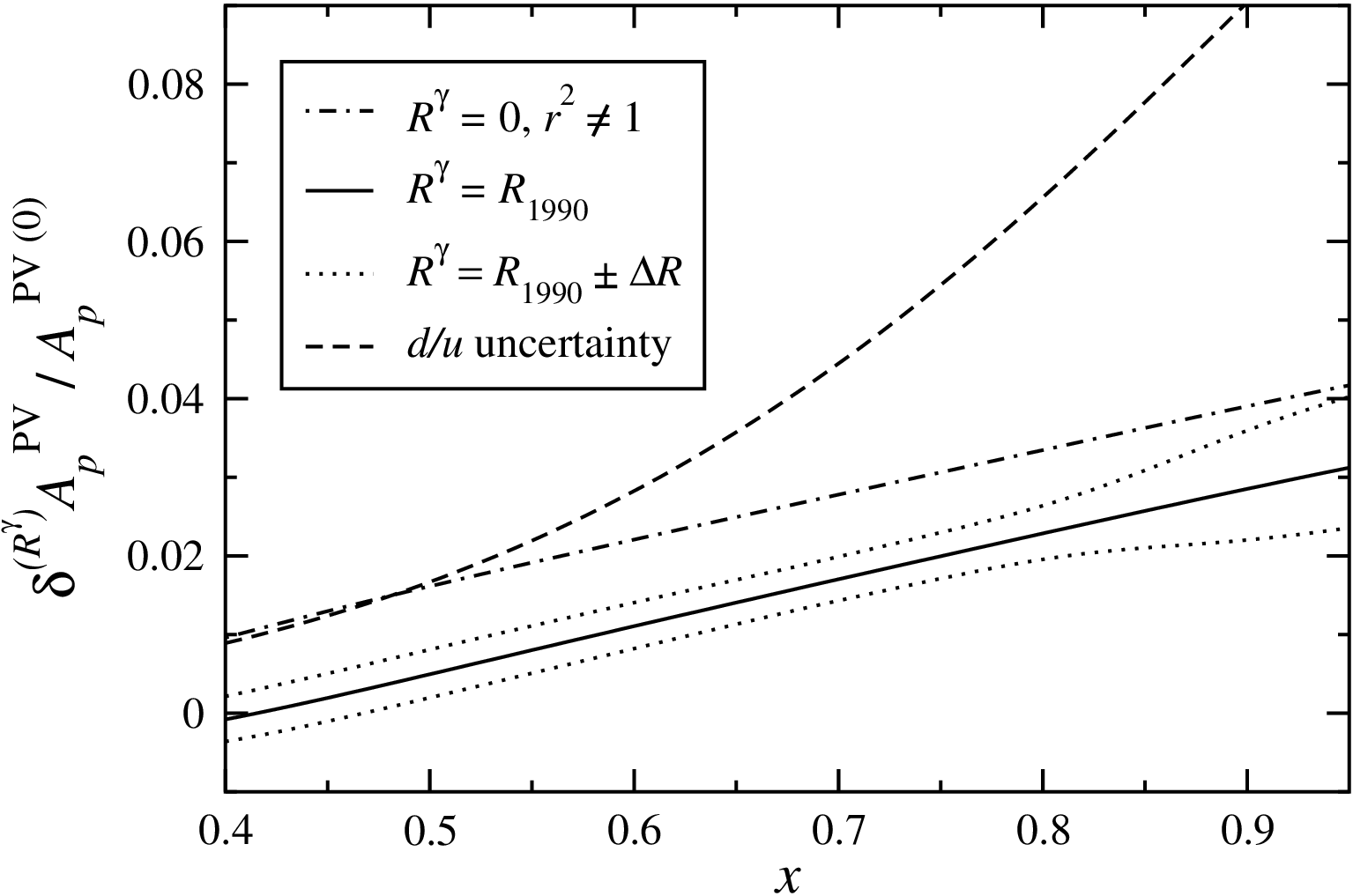}
\caption{Relative effects on the proton PV asymmetry $A^{\rm PV}_p$
	from the electromagnetic ratio $R^\gamma$ (keeping
	$R^{\gamma Z}=R^\gamma$), compared with the Bjorken limit
	asymmetry $A^{\rm PV\ (0)}_p$.
	The full results (solid), for $Q^2=5$~GeV$^2$, are compared
	with those for $R^\gamma=0$ (but $r^2 \neq 1$) (dot-dashed),
	with the dotted curves representing the uncertainty on
	$R^\gamma$ (from the ``$R_{1990}$'' parameterization of
	Ref.~\cite{R1990}).
	For reference the relative uncertainty in $A^{\rm PV}_p$
	arising from the $d/u$ ratio is also shown (dashed).}
\label{fig:dRg_Ap}
\end{figure}

At finite $Q^2$ the asymmetry $A^{\rm PV}_p$ depends not only on the
PDFs, but also on the longitudinal to transverse cross sections ratios
$R^\gamma$ and $R^{\gamma Z}$ for the electromagnetic and $\gamma Z$
interference contributions, respectively.
A number of measurements of the former have been taken at SLAC and
JLab \cite{R1990,R1998,RJLab}, and parametrizations of $R^\gamma$
in the DIS region exist.
In Fig.~\ref{fig:dRg_Ap} the relative effect on $A^{\rm PV}_p$
from $R^\gamma$ is shown via the ratio
$\delta^{(R^\gamma)} A^{\rm PV}_p / A^{\rm PV (0)}_p$,
where
$\delta^{(R^\gamma)} A^{\rm PV}_p
= A^{\rm PV}_p - A^{\rm PV (0)}_p$ is the difference between
the full asymmetry, with non-zero values of $R^\gamma$, and that
calculated in Bjorken limit kinematics, $A^{\rm PV (0)}_p$.

The effect on $A^{\rm PV}_p$ from the purely kinematical $r^2$
correction in the $Y_3$ term (with $R^\gamma=0$), compared with
the Bjorken limit prediction, is of the order $2-4\%$ over the
range $0.5 \lesssim x \lesssim 0.9$.
Including the $R^\gamma$ ratio from Ref.~\cite{R1990} reduces
the effect down to $\approx 1-3\%$ over the same range, with an
uncertainty of $\approx \pm 0.5\%$ for $x \lesssim 0.8$, and
$\approx 1\%$ at larger $x$.
This behavior can be easily understood from the expression for
$Y_3$ in Eq.~(\ref{eq:Y3}).
While the $r^2$ factor in the numerator of $Y_3$ leads to a larger
asymmetry at finite $Q^2$ (the $r^2$ dependence in the denominator
is in contrast diluted by the $y^2$ factor), a non-zero value for
$R^\gamma$ in the denominator of Eq.~(\ref{eq:Y3}) decreases $Y_3$
and lowers the overall correction.

These effects are to be compared with the relative change in
$A^{\rm PV}_p$ arising from different large-$x$ behaviors of the
$d/u$ ratio (dashed curve), expressed as a difference of the
asymmetries with the standard CTEQ6 \cite{CTEQ} PDFs and ones
with a modified $d/u$ ratio \cite{NP,Peng},
$\delta^{(d/u)} A_p^{\rm PV}/A_p^{\rm PV (0)}$.
where $A_p^{\rm PV (0)}$ is computed in terms of the standard
(unmodified) PDFs.
This is of the order 2\% for $x \sim 0.5$, but rises rapidly to
$\sim 10\%$ for $x \sim 0.9$.
While the kinematical and $R^\gamma$ corrections are smaller than
the (maximal) $d/u$ effect on the asymmetry, these must be included
in the data analysis in order to minimize the uncertainties on the
extracted $d/u$ ratio.

\begin{figure}[t]
\includegraphics[height=9cm]{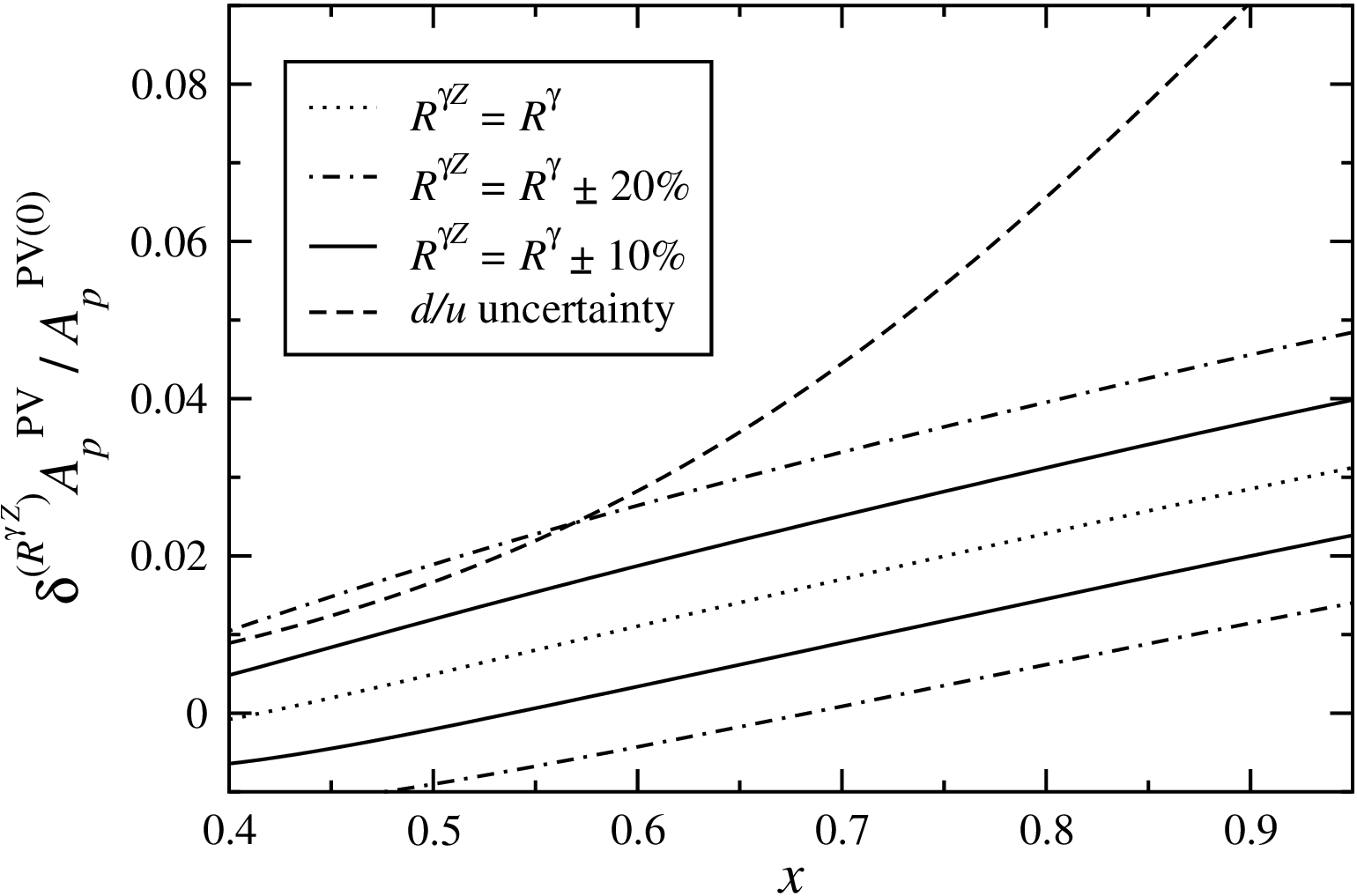}
\caption{Relative effects on the proton PV asymmetry $A^{\rm PV}_p$
	from the $\gamma Z$ interference ratio $R^{\gamma Z}$
	compared with the Bjorken limit asymmetry $A^{\rm PV\ (0)}_p$.
	The baseline result for $R^{\gamma Z} = R^\gamma$ (dotted)
	is compared with the effects of modifying $R^{\gamma Z}$ by
	$\pm 10\%$ (solid) and $\pm 20\%$ (dot-dashed),
	for $Q^2=5$~GeV$^2$.
	For reference the relative uncertainty
	$\delta^{(d/u)}A_p^{\rm PV}/A_p^{\rm PV (0)}$ from the
	$d/u$ ratio is also shown (dashed).}
\label{fig:dRgZ_Ap}
\end{figure}

In contrast to $R^\gamma$, no experimental information currently
exists on the interference ratio $R^{\gamma Z}$.
Since $R^{\gamma Z}$ enters in the relatively large $Y_1$ contribution
to $A^{\rm PV}_p$, any differences between $R^{\gamma Z}$ and
$R^\gamma$ could have important consequences for the asymmetry.
At high $Q^2$ one expects that $R^{\gamma Z} \approx R^\gamma$ at
leading twist, if the PVDIS process is dominated by single quark
scattering.
At low $Q^2$, however, since the current conservation constraints
are different for weak and electromagnetic probes, there may be
significant differences between these.

In Ref.~\cite{RlowQ} the ratios of $\sigma_L$ to $\sigma_T$ cross
sections for electromagnetic and weak processes were calculated using
a model which combines the low-$Q^2$ behavior from (axial) vector
meson dominance with perturbative QCD constraints at high $Q^2$
\cite{Alekhin}.
The resulting ratios $R^\gamma$ and $R^Z$ (which describes the
purely weak $Z$-exchange contribution) were found to differ by
($<1\%,\ 17\%,\ 22\%$) for $x = (0.4,\ 0.6,\ 0.85)$ at $Q^2 = 5$~GeV$^2$.
The differences at $Q^2 = 10$~GeV$^2$ were ($<1\%,\ 9\%,\ 23\%$) for
the same $x$ values \cite{SKpriv}.

For the interference ratio $R^{\gamma Z}$ one may expect qualitatively
similar behavior to that of $R^\gamma$ and $R^Z$, with $R^{\gamma Z}$
possibly lying in between the purely electromagnetic and weak ratios.
However, in the absence of a quantitative determination of $R^{\gamma Z}$, 
we take a more conservative estimate of the possible differences, and 
consider a range of possibilities, with $R^\gamma$ and $R^{\gamma Z}$
differing by 0\%, 10\% and 20\% for all $x$.

These are illustrated in Fig.~\ref{fig:dRgZ_Ap}, where we plot the ratio
$\delta^{(R^{\gamma Z})} A^{\rm PV}_p / A^{\rm PV (0)}_p$, with
$\delta^{(R^{\gamma Z})} A^{\rm PV}_p$ the difference between
the full asymmetry and that calculated in Bjorken limit kinematics,
$A^{\rm PV (0)}_p$.
The baseline correction with $R^{\gamma Z}=R^\gamma$ (dotted curve),
equivalent to the solid curve in Fig.~\ref{fig:dRg_Ap}),
with $R^\gamma$ from Ref.~\cite{R1990}, is compared with the effects
of modifying $R^{\gamma Z}$ by $\pm 10\%$ (solid) and $\pm 20\%$
(dot-dashed).
The result of such a modification, which comes through the $Y_1$ term
in the asymmetry, is an $\approx$ 1\% (2\%) shift of $A^{\rm PV}_p$
relative to the $R^{\gamma Z}$-independent asymmetry.
For $x \lesssim 0.6$, a 20\% difference between $R^{\gamma Z}$ and
$R^\gamma$ would be comparable to, or exceed, the maximal $d/u$
uncertainty considered here (dashed curve), although at larger $x$ the
sensitivity of $A^{\rm PV}_p$ to $d/u$ becomes increasingly larger.
As with the $R^\gamma$ corrections in Fig.~\ref{fig:dRg_Ap}, the
possible effects on the asymmetry due to $R^{\gamma Z}$ are potentially
significant, which warrants further work in understanding the
possible differences with $R^\gamma$ \cite{future}.

%%%%%%%%%%%%%%%%%%%%%%%%%%%%%%%%%%%%%%%%%%%%%%%%%%%%%%%%%%%%%%%%%%%%%%%%%
\section{PVDIS on the Deuteron}
\label{sec:d}

In the late 1970s parity-violating DIS on the deuteron provided an 
important early test of the standard model \cite{Prescott,Cahn}.
In the parton model, the asymmetry for an isoscalar deuteron becomes
independent of hadronic structure, and is given entirely by electroweak 
coupling constants.
At finite $Q^2$, however, contributions from longitudinal structure
functions, or from higher twist effects, may play a role.
The higher twists have been estimated in several of phenomenological 
model studies \cite{HT}.
More recently, it has been suggested that PVDIS on a deuteron could
also be sensitive to charge symmetry violation (CSV) effects in PDFs
(see Ref.~\cite{CSVreview} for a review of CSV in PDFs).
In this section we explore the contributions from kinematical
finite-$Q^2$ effects and the longitudinal structure functions on the
PV asymmetry, and assess their impact on the extraction of CSV effects.

% .......................................................................
\subsection{Finite-$Q^2$ Corrections}

Assuming the deuteron is composed of a proton and a neutron, and 
neglecting possible differences between free and bound nucleon PDFs,
the functions $a_1$ and $a_3$ in Eqs.~(\ref{eq:a1}) and (\ref{eq:a3})
for a deuteron target become:
\begin{subequations}
\begin{eqnarray}
\label{eq:a1d}
a_1^d &=& \frac{6}{5} \left( 2 C_{1u} - C_{1d} \right)\ ,	\\
\label{eq:a3d}
a_3^d &=& \frac{6}{5} \left( 2 C_{2u} - C_{2d} \right)\ .
\end{eqnarray}  
\end{subequations}
If in addition $R^\gamma_d \approx R^\gamma_p$ and
$R^{\gamma Z}_d \approx R^{\gamma Z}_p$, as is observed experimentally
\cite{R1990}, then the $y$-dependent terms in the deuteron asymmetry
become
$Y_1^d \approx Y_1^p \equiv Y_1$ and $Y_3^d \approx Y_3^p \equiv Y_3$.
The PV asymmetry can then be written as:
\begin{equation}
A^{\rm PV}
= -\left( \frac{3 G_F Q^2}{10 \sqrt{2}\pi\alpha} \right)
  \left[ Y_1 \left( 2 C_{1u} - C_{1d} \right)\
      +\ Y_3 \left( 2 C_{2u} - C_{2d} \right)\
  \right]\ ,
\end{equation}
which in the Bjorken limit ($Y_1 \to 1$, $Y_3 \to f(y)$) becomes
independent of hadron structure, and is a direct measure of the
electroweak coefficients $C_{iq}$.

\begin{figure}[t]
\includegraphics[height=9cm]{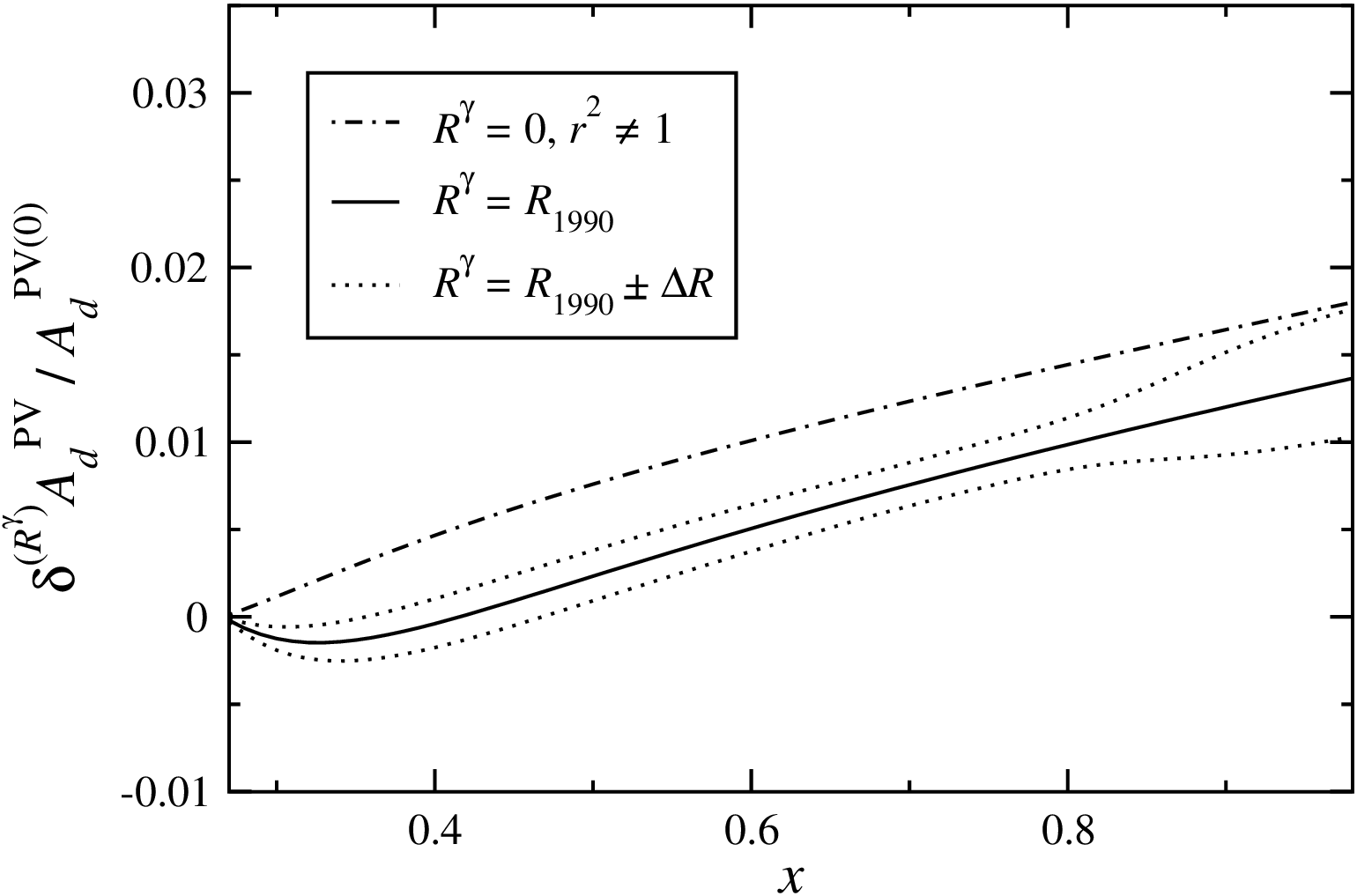}
\caption{Relative effects on the deuteron PV asymmetry
	$A^{\rm PV}_d$ from the electromagnetic ratio $R^\gamma$
	(with $R^{\gamma Z}=R^\gamma$), compared with the Bjorken
	limit asymmetry $A^{\rm PV\ (0)}_d$.
	The full results (solid), for $Q^2=5$~GeV$^2$, are compared
	with those for $R^\gamma=0$ (but $r^2 \neq 1$) (dot-dashed),
	with the dotted curves representing the uncertainty on
	$R^\gamma$ (from the ``$R_{1990}$'' parameterization of
	Ref.~\cite{R1990}).}
\label{fig:dRg_Ad}
\end{figure}

\begin{figure}[t]
\includegraphics[height=9cm]{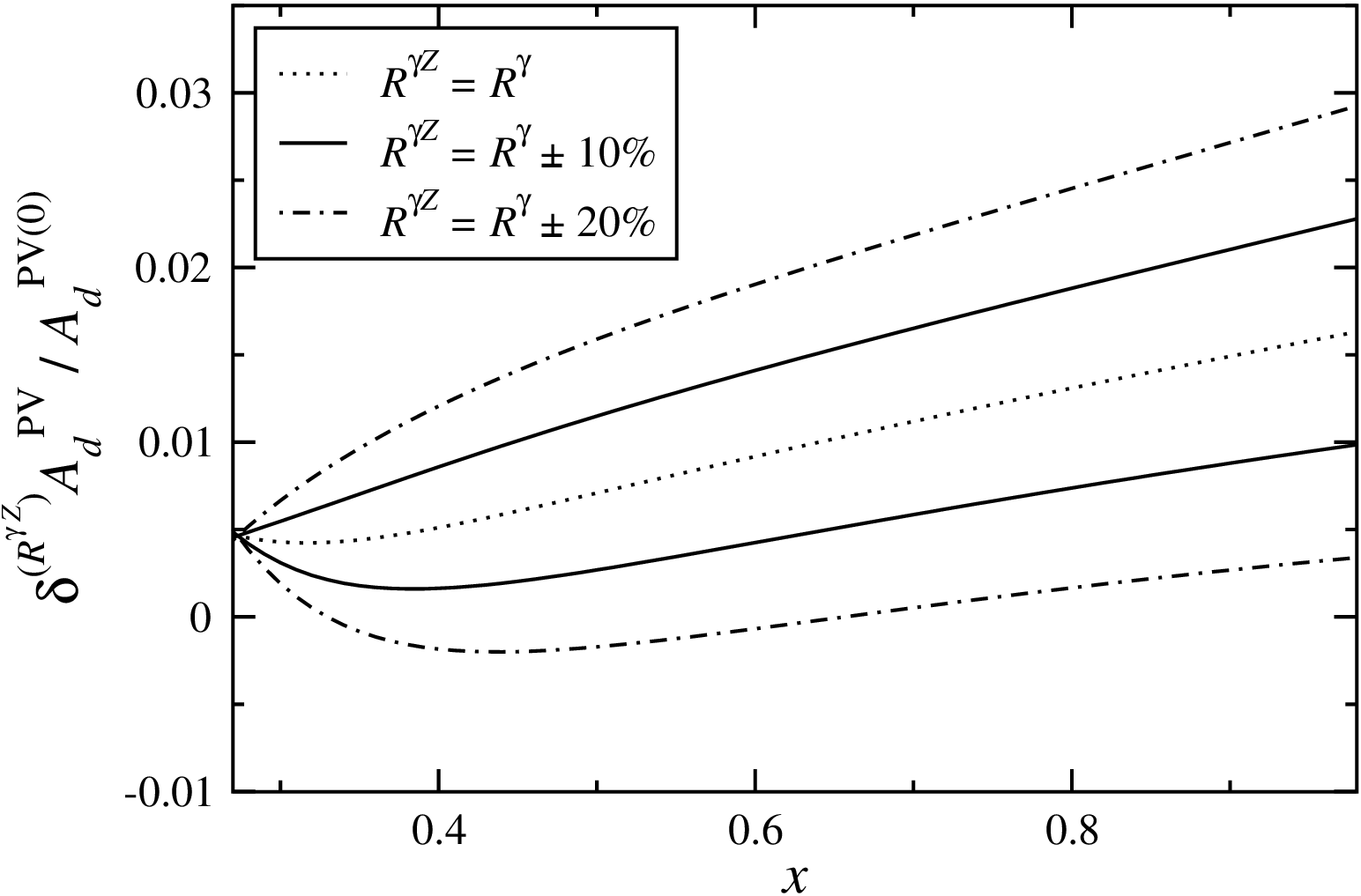}
\caption{Relative effects on the deuteron PV asymmetry $A^{\rm PV}_d$
	from the $\gamma Z$ interference ratio $R^{\gamma Z}$
	compared with the Bjorken limit asymmetry $A^{\rm PV\ (0)}_d$.
	The baseline result for $R^{\gamma Z} = R^\gamma$ (dotted)
	is compared with the effects of modifying $R^{\gamma Z}$ by
	$\pm 10\%$ (solid) and $\pm 20\%$ (dot-dashed),
	for $Q^2=5$~GeV$^2$.}
\label{fig:dRgZ_Ad}
\end{figure}

In Fig.~\ref{fig:dRg_Ad} the relative effect on $A^{\rm PV}_d$
from $R^\gamma$ is shown via the ratio
$\delta^{(R^\gamma)} A^{\rm PV}_d / A^{\rm PV (0)}_d$,
where
$\delta^{(R^\gamma)} A^{\rm PV}_d$ is the difference between the
full asymmetry and that calculated in Bjorken limit kinematics, 
$A^{\rm PV (0)}_d$.
The correction due to $R^\gamma$ is qualitatively similar to that
for the proton asymmetry in Fig.~\ref{fig:dRg_Ap}, although slightly
smaller.
The effect on $A^{\rm PV}_d$ from the purely kinematical $r^2$ correction
in the $Y_3$ term (with $R^\gamma=0$) is an increase of order $1\%$ over
the Bjorken limit asymmetry in the range $0.5 \lesssim x \lesssim 0.9$.
Inclusion of the $R^\gamma$ ratio cancels the correction somewhat,
reducing it to $\lesssim 0 - 0.5\%$ for $x \lesssim 0.6$, and to
$\lesssim 0.5 - 1\%$ for $x > 0.6$.

The effects of a possible difference between $R^{\gamma Z}$ and
$R^\gamma$ are illustrated in Fig.~\ref{fig:dRgZ_Ad} through the
ratio 
$\delta^{(R^{\gamma Z})} A^{\rm PV}_d / A^{\rm PV (0)}_d$,
where
$\delta^{(R^{\gamma Z})} A^{\rm PV}_p$ is the difference between   
the full and Bjorken limit asymmetries.
As for the proton in Fig.~\ref{fig:dRgZ_Ap}, the baseline correction
with $R^{\gamma Z}=R^\gamma$ (dotted curve) is compared with the effects
of modifying $R^{\gamma Z}$ by a constant $\pm 10\%$ (solid) and
$\pm 20\%$ (dot-dashed).
This conservative range is, as for the proton, motivated by the
phenomenological study of $R^\gamma$ and $R^Z$ in Ref.~\cite{RlowQ},
and the relatively weak isospin dependence of $R^\gamma$ 
\cite{R1990,R1998}.
This results in an additional $\approx$ 0.5\% (1\%) shift of
$A^{\rm PV}_d$ for a 10\% (20\%) modification relative to the
baseline asymmetry for $x > 0.5$.
Such effects will need to be accounted for if one wishes to compare
with the standard model predictions, or when extracting CSV effects
in PDFs, which we discuss in the next section.

% .......................................................................
\subsection{Charge Symmetry Violation}

In the entire discussion above an implicit assumption has been made
that charge symmetry is exact, namely that the quark distributions
in the proton and neutron are related by $u^p = d^n$ and $u^n = d^p$.
Quark mass differences and electromagnetic effects are expected,
however, to give rise to (small) violations of charge symmetry in
PDFs, which may be parameterized by:
\begin{subequations}
\begin{eqnarray}
\delta u &=& u^p - d^n\ , \\
\delta d &=& d^p - u^n\ .
\end{eqnarray}
\end{subequations}
Non-zero values of $\delta u$ and $\delta d$ have been predicted in
nonperturbative models of the nucleon \cite{CSVmodels}, and can in
addition arise from radiative QED effects in $Q^2$ evolution 
\cite{MRSTCSV,MRSTQED,GJRQED}.

\begin{figure}[t]
\includegraphics[height=9cm]{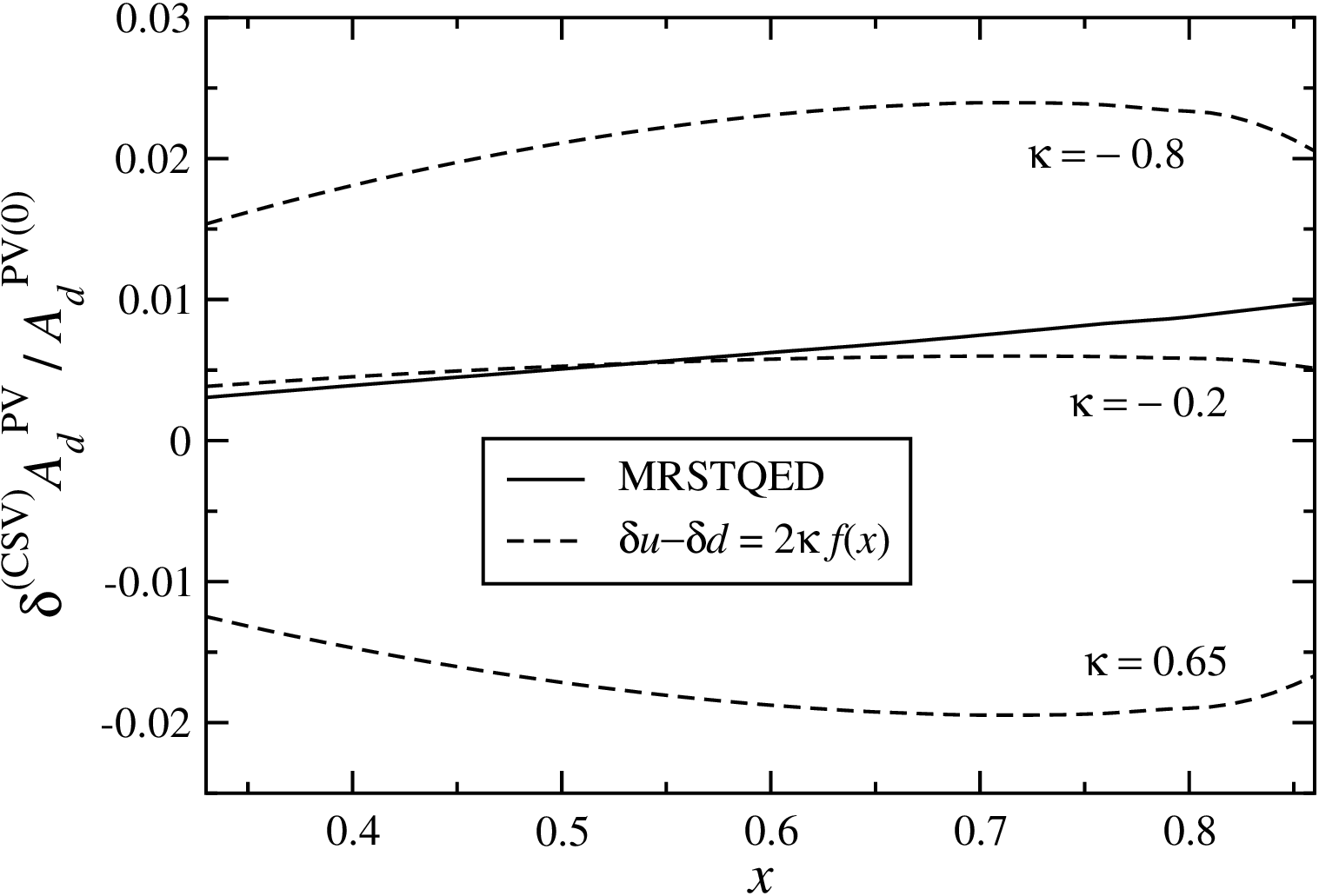}
\caption{Relative effects on the deuteron PV asymmetry $A^{\rm PV}_d$
	of CSV in PDFs, compared with the charge symmetric asymmetry.
	The CSV distributions $\delta u - \delta d$ are from the
	MRSTQED fit \cite{MRSTQED} (solid) and from the parametrization
	$\delta u - \delta d = 2 \kappa f(x)$ (dashed, see text),
	with $\kappa = -0.2$ (best fit), and the two 90\% confidence
	levels, $\kappa = -0.8$ and $\kappa = +0.65$ \cite{MRSTCSV}.}
\label{fig:dCSV}
\end{figure}

It is convenient to define the $u$ and $d$ quark distributions
in the presence of CSV according to \cite{KK}:
\begin{subequations}
\begin{eqnarray}
u\ \equiv\ u^p - \frac{\delta u}{2} &=& d^n + \frac{\delta u}{2}\ , \\
d\ \equiv\ d^p - \frac{\delta d}{2} &=& u^n + \frac{\delta d}{2}\ . 
\end{eqnarray}
\end{subequations}
With these definitions, the deuteron functions $a_1^d$ and $a_3^d$
in the $A_d^{\rm PV}$ asymmetry can be written:
\begin{subequations}
\begin{eqnarray}
a_1^d &=& a_1^{d(0)} + \delta^{\rm (CSV)} a_1^d\ ,	\\
a_3^d &=& a_3^{d(0)} + \delta^{\rm (CSV)} a_3^d\ ,
\end{eqnarray}
\end{subequations}
where $a_1^{d(0)}$ and $a_3^{d(0)}$ are given by Eqs.~(\ref{eq:a1d})
and (\ref{eq:a3d}), respectively.
The fractional CSV corrections are given by:
\begin{subequations}
\begin{eqnarray}
\label{eq:csva1}
\frac{\delta^{\rm (CSV)} a_1^d}{a_1^{d(0)}}
&=& \left( -\frac{3}{10} + \frac{2 C_{1u} + C_{1d}}{2 (2 C_{1u} - C_{1d})}
    \right)
    \left( \frac{\delta u - \delta d}{u + d} \right)\ ,	\\
\frac{\delta^{\rm (CSV)} a_3^d}{a_3^{d(0)}}
&=& \left( -\frac{3}{10} + \frac{2 C_{2u} + C_{2d}}{2 (2 C_{2u} - C_{2d})}
    \right)
    \left( \frac{\delta u - \delta d}{u + d} \right)\ .
\label{eq:csva3}
\end{eqnarray}
\end{subequations}

\begin{figure}[t]
\includegraphics[height=9cm]{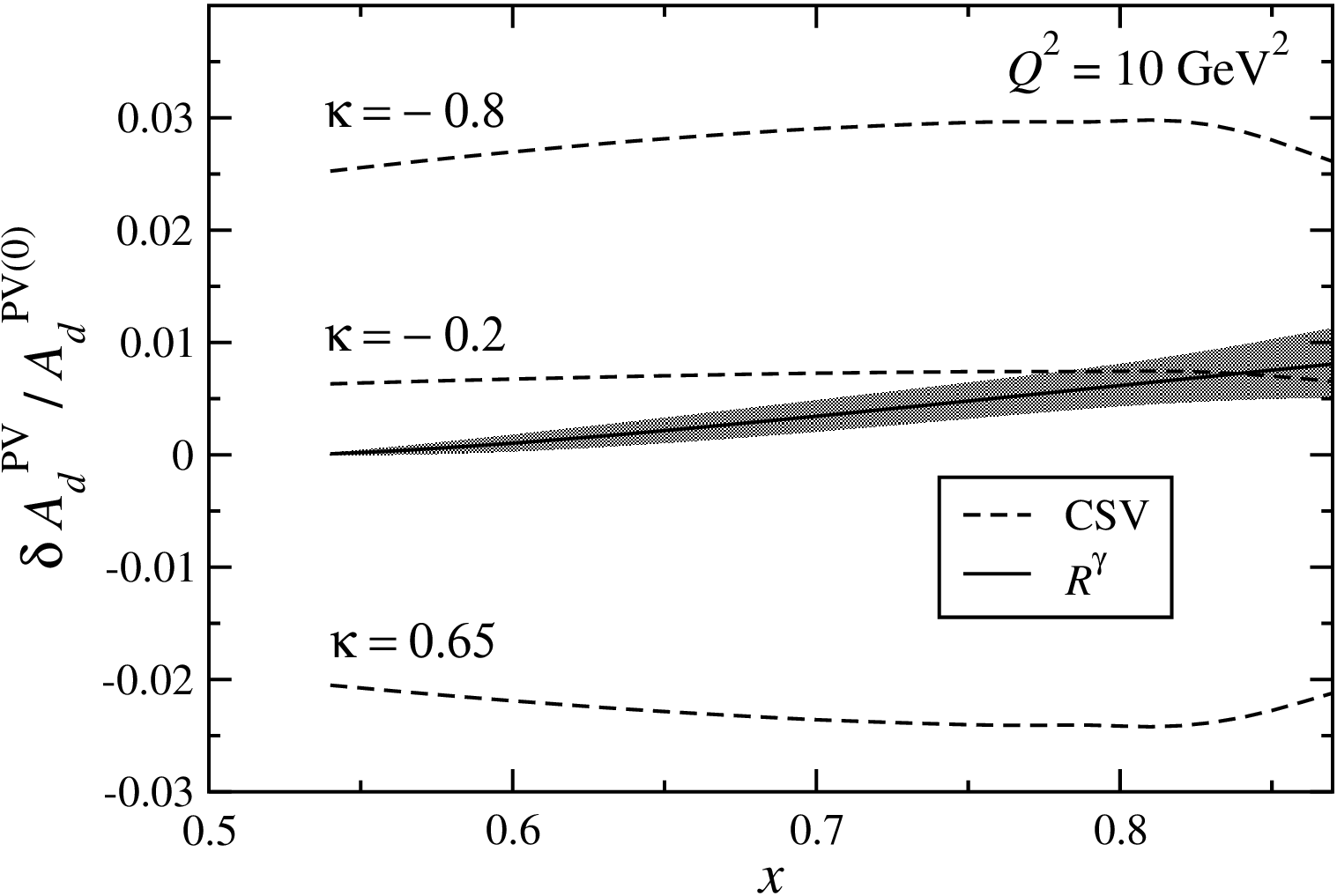}
\caption{Relative effects on the deuteron PV asymmetry $A^{\rm PV}_d$
	from CSV in PDFs \cite{MRSTCSV} (dashed, see Fig.~\ref{fig:dCSV})
	and from $R^\gamma$ \cite{R1990} (with $R^{\gamma Z}=R^\gamma$,
	solid) at $Q^2 = 10$~GeV$^2$, compared with the charge symmetric
	asymmetry in Bjorken limit kinematics.
	The shaded area represents the uncertainty in $R^\gamma$.}
\label{fig:dRg_AdQ10}
\end{figure}

In Fig.~\ref{fig:dCSV} we plot the effect of CSV in valence PDFs on
the deuteron asymmetry $A_d^{\rm PV}$.
The asymmetry using the MRSTQED parametrization \cite{MRSTQED} of
$\delta u - \delta d$ (solid curve) gives an $\approx 0.5 - 1\%$ effect
for $0.5 \lesssim x \lesssim 0.9$, similar to the effect predicted
from nonperturbative (bag model) calculations \cite{CSVmodels}.
The phenomenological fit \cite{MRSTCSV}
$\delta u - \delta d = 2 \kappa f(x)$, with
$f(x) = x^{-1/2} (1-x)^4 (x-0.0909)$
and $\kappa$ a free parameter, results in a similar correction to the 
asymmetry, $\sim 0.5\%$ for most of the $x$ range considered.
The best fit gives $\kappa = -0.2$, although the constraints on
$\kappa$ are relatively weak, with values of $\kappa = -0.8$ and
$+0.65$ giving $\sim 1.5 - 2\%$ effect for $0.5 \lesssim x \lesssim 0.8$
at the 90\% confidence level.

For the central values (best fit parameters), the magnitude of the CSV
effect on the asymmetry at $Q^2 = 5$~GeV$^2$ is similar to that due to
the finite-$Q^2$ kinematics ($r^2 \neq 1$, $R^\gamma \neq 0$) seen in
Fig.~\ref{fig:dRg_Ad}, and may be smaller than that due to possible
differences between $R^{\gamma Z}$ and $R^\gamma$ in 
Fig.~\ref{fig:dRgZ_Ad}.
Unless the finite-$Q^2$ corrections are known to greater accuracy than
at present, they may impede the unambiguous extraction of CSV effects
from the asymmetry.

On the other hand, since the finite-$Q^2$ corrections are expected to
decrease with $Q^2$, while the CSV effects are leading twist effects,
a cleaner separation should be possible at larger $Q^2$.
In Fig.~\ref{fig:dRg_AdQ10} the effect of $R^\gamma$ on $A^{\rm PV}_d$
(solid) is compared with the CSV results \cite{MRSTCSV} for different
$\kappa$ values (dashed) at $Q^2 = 10$~GeV$^2$.
The deviation from the Bjorken limit kinematics of the
$\delta^{(R^\gamma)}$ curve is clearly less than the corresponding
result at $Q^2 = 5$~GeV$^2$ in Fig.~\ref{fig:dRg_Ad} (the shaded region
here indicates the uncertainty in $R^\gamma$), whereas the CSV results
are similar to those at the lower $Q^2$.
The contrast is especially striking at $x \sim 0.6$, where the CSV
effects are several times larger than the correction to $A^{\rm PV}_d$
due to $R^\gamma$.
At larger $x$ the CSV effects for the central $\kappa$ value become
comparable to the $R^\gamma$ uncertainty, however, the 90\% confidence
level corrections ($\kappa = -0.8$ and +0.65) are of the order 2\%
and are still several times larger than the $R^\gamma$ uncertainty.

These results suggest that if the CSV effects in PVDIS from the
deuteron are of the order $\sim 0.5\%$, the optimal value of $x$
to observe them would be $x \sim 0.6$ at $Q^2=10$~GeV$^2$.
If the CSV effects are of order $\sim 2\%$, they should be clearly
visible over a larger $x$ range, even up to $x \approx 0.8$.
Note that the minimum value of $x$ attainable at the $Q^2=10$~GeV$^2$
kinematics ($x \approx 0.53$) is somewhat smaller than at the lower
$Q^2$ vales because at fixed incident energy and $Q^2$ the fractional
lepton energy loss exceeds unity at higher $x$.

%%%%%%%%%%%%%%%%%%%%%%%%%%%%%%%%%%%%%%%%%%%%%%%%%%%%%%%%%%%%%%%%%%%%%%%%%
\section{Prospects for PVDIS on Polarized Hadrons}
\label{sec:pol}

In this section we explore the possibility of extracting
{\em spin-dependent} PDFs in parity-violating unpolarized-electron 
scattering from a {\em polarized} hadron.
In particular, we examine the sensitivity of the polarized proton,
neutron and deuteron PVDIS asymmetries to the polarized $\Delta u$
and $\Delta d$ distributions at large $x$, where these are poorly
known.
The $\Delta d$ distribution in particular remains essentially
unknown beyond $x \approx 0.6$.

The PV differential cross-section (with respect to the variables
$x$ and $y$) for unpolarized electrons on longitudinally polarized
nucleons can generally be written in terms of 5 spin-dependent
structure functions \cite{Ans}:
\begin{eqnarray}
\frac{d^2 \sigma^{\rm PV}}{dxdy}(\bar{\lambda}, S_L)
&=& 2x \left( 2-y-\frac{xyM}{E} \right) g_1^{\gamma Z}\
 -\ \frac{4x^2M}{E}\ g_2^{\gamma Z}\
 +\ \frac{2}{y} \left( 1-y-\frac{xyM}{2E} \right) g_3^{\gamma Z}
	\nonumber\\
& &
 -\ \frac{2}{y} \left( 1+\frac{xM}{E} \right)
	        \left( 1-y-\frac{xyM}{2E} \right) g_4^{\gamma Z}\
 +\ 2xy \left( 1+\frac{xM}{E} \right) g_5^{\gamma Z}\ ,
\end{eqnarray}
where the nucleon (longitudinal) spin vector $S_L$ is given by 
$S^{\mu}_L = (0;0,0,1)$, and $\bar{\lambda}$ is the average over 
$\lambda = +1$ and $\lambda = -1$ (see Eq.~(\ref{eq:LgZ}).
The analog of the PV asymmetry in Eq.~(\ref{eq:APVdef}) for a
polarized target can be defined as:
\begin{equation}
\Delta A^{\rm PV}
= \frac{\sigma^{\rm PV} (\bar{\lambda}, S_L)
      - \sigma^{\rm PV} (\bar{\lambda}, -S_L)}
       {\sigma^{\rm PV} (\bar{\lambda}, S_L)
      + \sigma^{\rm PV} (\bar{\lambda}, -S_L)}\ ,
\end{equation}
where $\sigma^{\rm PV} (\bar{\lambda}, S_L)
\equiv d^2 \sigma^{\rm PV}/dxdy$.
Some of the structure functions $g_{1-5}^{\gamma Z}$ have simple
parton model interpretations, while others do not.
At present there is no phenomenological information about these
structure functions.
In order to proceed, we shall therefore consider the asymmetry in
the high energy limit, $M/E \to 0$, which eliminates the structure
function $g_2^{\gamma Z}$.
In this limit, the operator product expansion gives rise to the
relation $g_3^{\gamma Z} - g_4^{\gamma Z} = 2xg_5^{\gamma Z}$,
which further eliminates one of the functions.
Furthermore, in the parton model the $g_4^{\gamma Z}$ structure
function vanishes, leaving the Callan-Gross-like relation
$g_3^{\gamma Z} = 2x g_5^{\gamma Z}$.
In terms of the remaining two structure functions, the spin-dependent
PV asymmetry can be written:
\begin{equation} 
\Delta A^{\rm PV}
= \frac{G_F Q^2}{4 \sqrt{2}\pi\alpha}
  \left( g_A^e\ f(y)\ \frac{g_1^{\gamma Z}}{F_1^{\gamma}}\
      +\ g_V^e\ \frac{g_5^{\gamma Z}}{F_1^{\gamma}}
  \right)\ ,
\end{equation}  
where the kinematical factor $f(y)$ is given in Eq.~(\ref{eq:Y3bj}).

\begin{figure}[t]
\includegraphics[height=9cm]{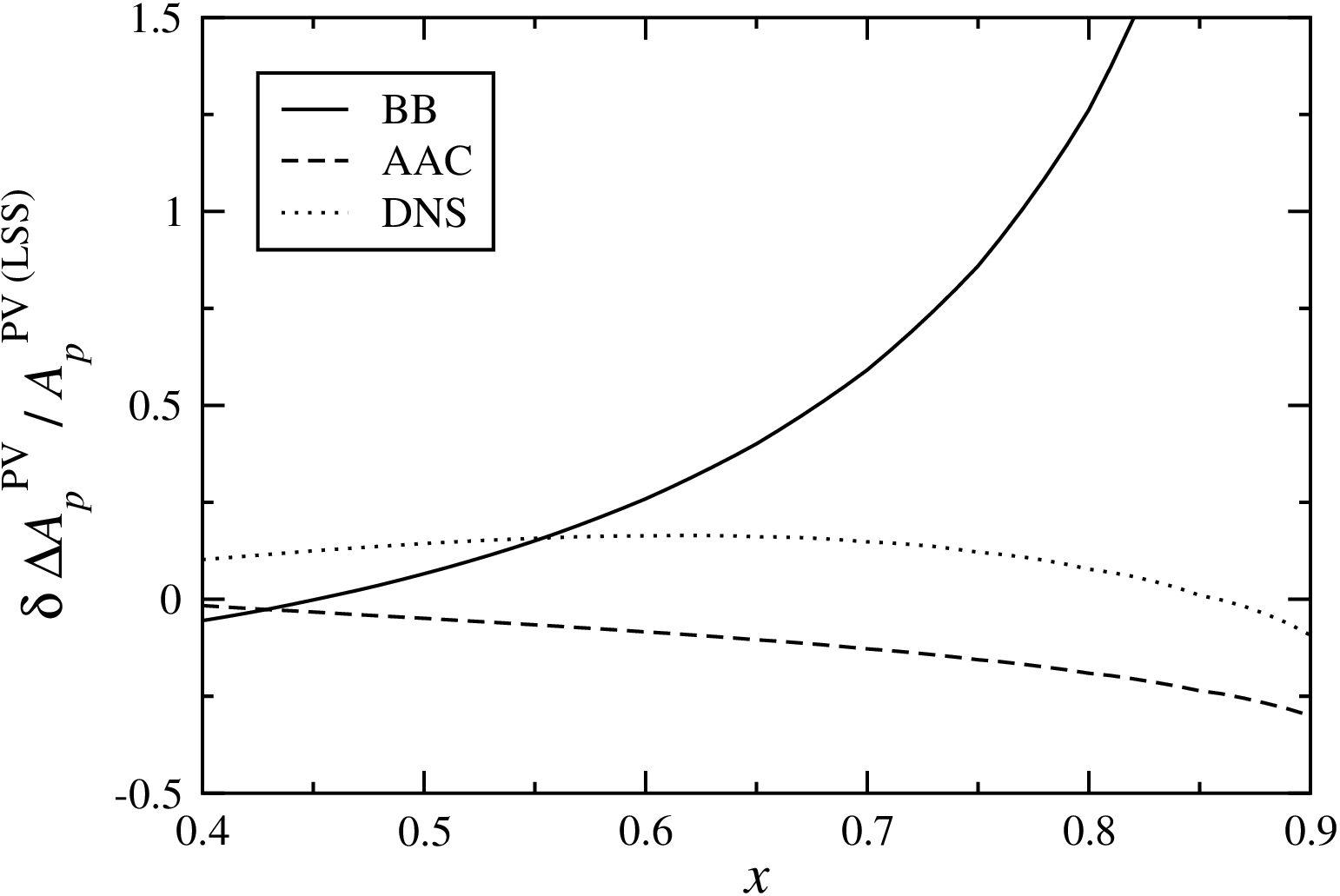}
\caption{Sensitivity of the polarized proton PV asymmetry
	$\Delta A^{\rm PV}_p$ on the spin-dependent
	$\Delta u$ and $\Delta d$ distributions.
	The asymmetries for the BB \cite{BB} (solid), AAC \cite{AAC}
	(dashed) and DNS \cite{DNS} (dotted) distributions are
	evaluated relative to the baseline asymmetry for the LSS
	PDFs \cite{LSS}.}
\label{fig:dDA}
\end{figure}

In the QCD parton model the $g_1^{\gamma Z}$ and $g_5^{\gamma Z}$
structure functions can be expressed in terms of helicity dependent
PDFs $\Delta q$ as \cite{Ans}:
\begin{subequations}
\begin{eqnarray}
g_1^{\gamma Z}
&=& \sum_q e_q\ g_V^q
    \left( \Delta q + \Delta \bar{q} \right)\ , \\
g_5^{\gamma Z}
&=& \sum_q e_q\ g_A^q\
    \left( \Delta q - \Delta \bar{q} \right)\ ,
\end{eqnarray}
\end{subequations}
where $\Delta q$ is a function of $x$ and $Q^2$.
Using these expressions, the PV asymmetries for proton, neutron and 
deuteron (which in this analysis we take to be a sum of proton and 
neutron) targets can then be written \cite{KK}:
\begin{subequations}
\begin{eqnarray}
\Delta A^{\rm PV}_p
&=& \frac{6\ G_F Q^2}{4 \sqrt{2}\pi\alpha}
    \left[ (2 C_{1u} \Delta u - C_{1d} \Delta d) f(y)
	 + (2 C_{2u} \Delta u - C_{2d} \Delta d)
    \right]
    \left( {1 \over 4 u + d} \right)\ ,			\\
\Delta A^{\rm PV}_n
&=& \frac{6\ G_F Q^2}{4 \sqrt{2}\pi\alpha}
    \left[ (2 C_{1u} \Delta d - C_{1d} \Delta u) f(y)
	 + (2 C_{2u} \Delta d - C_{2d} \Delta u)
    \right]
    \left( {1 \over u + 4 d} \right)\ ,			\\
\Delta A^{\rm PV}_d
&=& \frac{3\ G_F Q^2}{10 \sqrt{2}\pi\alpha}
    \left[ (2 C_{1u} - C_{1d}) f(y)
         +  2 C_{2u} - C_{2d}
    \right]
    \left( {\Delta u + \Delta d \over u + d} \right)\ .
\end{eqnarray}
\end{subequations}
Generalizations to higher order are straightforward, however, just as in 
the unpolarized case, care should be taken with large-$x$ resummations
\cite{resum}, which could modify some of the quantitative conclusions
at $x \sim 1$.

In Fig.~\ref{fig:dDA} we illustrate the sensitivity of the proton
asymmetry $\Delta A^{\rm PV}_p$ to the $\Delta u$ and $\Delta d$ PDFs,
by comparing the difference $\delta \Delta A^{\rm PV}_p$ in the
asymmetry arising from different parametrizations \cite{BB,AAC,DNS},
relative to the LSS parametrization \cite{LSS}.
The effects at intermediate $x$, $x \sim 0.5 - 0.6$, are of order 20\%,
however, these increase rapidly with $x$.
At $x \approx 0.7 - 0.8$ the AAC \cite{AAC}, DNS \cite{DNS} and LSS
\cite{LSS} parametrizations give asymmetries that are within
$\sim 20\%$ of each other, whereas the BB fit \cite{BB} deviates
by $50 - 100\%$ in this range.
The results for neutron and deuteron targets are found to be very
similar to those in Fig.~\ref{fig:dDA}.
While this does not constitute a systematic error on the uncertainty
in $\Delta A^{\rm PV}_p$ due to PDFs, it does indicate the sensitivity
of polarized PVDIS to helicity distributions at large $x$, and suggests
that a measurement of $\Delta A^{\rm PV}_p$ at the $10 - 20\%$ level
could discriminate between different PDF behaviors.

% Finally, for completeness it should be possible in principle to extract 
% data on the PDF quantities $\frac{\Delta u}{u}$ and $\frac{\Delta d}{d}$ 
% from these polarized asymmetries given our knowledge of unpolarized
% PDF behavior and experimental value of the DIS couplings.
%
% We find:
%
% \begin{subequations}
% \begin{eqnarray}
% \frac{\Delta u}{u} 
% &=& \frac{\sqrt{8} \pi \alpha}{3G_F Q^2} \cdot
%     \left( 2 (C_{1u} f(y) + C_{2u}) (4 + d/u) \Delta A^p 
%         + (C_{1d}+C_{2d})(4 d/u + 1) \Delta A^n
%     \right)
%     \frac{1}{4 [C_{1u}f(y)+C_{2u}]^2 + [C_{1d}f(y)+C_{2d}]^2}\ ,  \\
%
% \frac{\Delta d}{d} 
% &=& \frac{\sqrt{8} \pi \alpha}{3G_F Q^2} \cdot 
%     \frac{[2(C_{1u}f(y) + C_{2u})(4+u/d) \Delta A^n 
%         + (C_{1d}+C_{2d})(4 u/d+1)\Delta A^p]}
%        {4[C_{1u}f(y)+C_{2u}]^2 + [C_{1d}f(y)+C_{2d}]^2}\ .
% \end{eqnarray}
% \end{subequations}
%
% Alternatively, we may write these in terms of the proton and deuteron
% asymmetries:
%
% \begin{subequations}
% \begin{eqnarray}
% \frac{\Delta u}{u} 
% &=& \frac{\sqrt{8}\pi\alpha
%   [(4+d/u)C_1(y)\Delta A^p + 5(1+d/u)[C_{1d}f(y)+C_{2d}]\Delta A^d]}
%   {3G_F Q^2 (C_1(y) \cdot C_2(y)}\ ,    \\
%
% \frac{\Delta d}{d} 
% &=& \frac{\sqrt{8}\pi\alpha [\frac{5(u/d+1)\Delta A^d}{C_1(y)} 
%         - \frac{(4u/d+1)\Delta A^p}{C_{1u}f(y)+C_{2u}}}      
%        {1+\frac{C_{1d}f(y)+C_{2d}}{2(C_{1u}f(y)+C_{2u}}}\ .
% \end{eqnarray}
% \end{subequations}
%
% $\bullet$
% Write $\Delta A^{\rm PV}$ in terms of $\Delta q$ distributions.

%%%%%%%%%%%%%%%%%%%%%%%%%%%%%%%%%%%%%%%%%%%%%%%%%%%%%%%%%%%%%%%%%%%%%%%%%
\section{Conclusions}
\label{sec:conc}

Parity-violating deep inelastic scattering provides a unique tool with
which to study novel aspects of the partonic structure of the nucleon,
such as the flavor dependence of PDFs in the region $x \sim 1$ or 
charge-symmetry violation in PDFs, or even more exotic physics beyond
the standard model.
In this paper we have examined the sensitivity of the PVDIS process to
finite-$Q^2$ effects which can give rise to important corrections to
parton model results at scales $Q^2 \sim$ few GeV$^2$.

The suppression of the leptonic vector couplings $C_{2q}$ relative
to the axial-vector couplings $C_{1q}$ leads to the dominance of the
parity-violating asymmetry $A^{\rm PV}$ by the hadronic-vector term
$a_1$. 
In practice the hadronic $a_3$ axial-vector contribution amounts to
some some 20\% of the total, for both proton and deuteron targets,
and must be accounted for in quantitative numerical analyses.
In particular, the $a_3$ term is associated with the kinematical
dependence on the ratio $R^\gamma$ of electromagnetic longitudinal
to transverse photon cross sections.

For the proton asymmetry, which is sensitive to the $d/u$ parton
distribution function ratio at large $x$ \cite{Souder}, the corrections
from non-zero values of $r^2-1 = 4 M^2 x^2/Q^2$ and $R^\gamma$
lead to an $\approx 1 - 2\%$ shift in $A^{\rm PV}_p$ over the range
$0.6 \lesssim x \lesssim 0.8$, with an uncertainty of $\pm 0.5\%$,
increasing to an $\approx 3\%$ shift for $x \approx 0.9$ with an
uncertainty of $\pm 1\%$.
This is to be compared with a sensitivity ranging from
$\approx 3 - 10\%$ in the asymmetry due to different behaviors
of the $d/u$ ratio for the same range of $x$.

The correction from the longitudinal to transverse $\gamma-Z$
interference cross section ratio $R^{\gamma Z}$, which has an
unexplored phenomenology, could contribute to $A^{PV}_p$ if it
differs significantly from $R^{\gamma}$, especially given that
$R^{\gamma Z}$ enters through the large, $C_{1q}$-weighted vector
term.
While we expect that $R^{\gamma Z} \approx R^\gamma$ at high $Q^2$,
deviations of 10\% (20\%) at $Q^2 \sim 5$~GeV$^2$ would result in
$\approx 1\% (2\%)$ shift in the asymmetry.
For $x \lesssim 0.6$ this would be comparable to the maximal $d/u$
effect on $A^{PV}_p$, although at larger $x$ the sensitivity to
$d/u$ becomes increasingly larger.

For the deuteron asymmetry, the low-$Q^2$ corrections due to $r^2$
and $R^\gamma$ are similar to those for the proton at $Q^2 = 5$~GeV$^2$,
although slightly smaller, and lead to an increase of $\lesssim 1\%$
in the Bjorken-limit asymmetry for $x \lesssim 0.85$.
Possible deviations of $R^{\gamma Z}$ from $R^\gamma$ can lead to
further corrections to $A^{PV}_d$, ranging from $\approx 0.5 - 1\%$
for $10 - 20\%$ differences between the ratios.
Such effects are comparable with those arising from charge symmetry
violation in PDFs, as estimated in nonperturbative models and
phenomenological fits to data.
This suggests that without better knowledge of the low-$Q^2$
corrections, and $R^{\gamma Z}$ in particular, extracting unambiguous
information on CSV at these kinematics may be difficult.
On the other hand, since the CSV effects on PDFs are leading twist,
they will persist at larger $Q^2$ where the corrections due to
$R^{\gamma (\gamma Z)}$ will be suppressed.
A cleaner separation of the CSV effects should therefore be more
feasible at larger $Q^2$, $Q^2 \approx 10$~GeV$^2$, where the CSV
effects can be up to several times larger than those due to finite
longitudinal cross sections.
The proposed experiments at JLab with 12~GeV \cite{JLab12} plan
to measure the PV asymmetries over a wide range of $Q^2$ and $y$
at fixed values of $x$, which should enable the various effects
to be disentangled.

Finally, we have explored the possibility of constraining spin-dependent
PDFs from PVDIS of unpolarized leptons from polarized hadrons.
Currently there is considerable uncertainty in the behavior of
the $\Delta u$ and $\Delta d$ helicity distributions at large $x$,
and our estimates suggest that, while challenging, measurement of
polarized PV asymmetries at the $10 - 20\%$ level could discriminate
between different PDF behaviors for $x > 0.7$.
Whether this can be achieved experimentally in the foreseeable
future remains to be seen \cite{XiaoPol}.

For the future, a number of outstanding issues can be identified.
Firstly, the present exploratory studies need to be complemented by
more quantitative determinations of $R^{\gamma Z}$, either from
model calculations or from phenomenology, in order to reduce the
uncertainties in the low-$Q^2$ corrections to the asymmetries.
In addition, target mass corrections to the interference structure
functions $F_{1-3}^{\gamma Z}$ should be computed, which may have
important consequences in the large-$x$, low-$Q^2$ region \cite{TMC}.
Furthermore, the effects of higher twist contributions to electroweak
structure functions must be taken into account; although these have been
estimated in nonperturbative models to be relatively small \cite{HT},
they nonetheless need to be included in a complete analysis of PVDIS
at few-GeV$^2$ scales.
The work presented here sets the stage for more detailed theoretical
analysis \cite{future} in the run-up to future precision PVDIS
measurements at facilities such as Jefferson Lab \cite{JLab6,JLab12}.

%%%%%%%%%%%%%%%%%%%%%%%%%%%%%%%%%%%%%%%%%%%%%%%%%%%%%%%%%%%%%%%%%%%%%%%%%
\section*{Acknowledgements}

We thank K.~S.~Kumar, J.~T.~Londergan, K.~Paschke, P.~Reimer, P.~Souder
and X.~Zheng for helpful discussions and communications, and 
S.~A.~Kulagin for sending the results of Ref.~\cite{RlowQ}.
T.~H. thanks U.~S. Department of Energy's Science Undergraduate
Laboratory Internships (SULI) Program at Jefferson Lab and the
Jefferson Lab Theory Center for support.
This work was supported by the DOE contract No. DE-AC05-06OR23177,
under which Jefferson Science Associates, LLC operates Jefferson Lab.

%%%%%%%%%%%%%%%%%%%%%%%%%%%%%%%%%%%%%%%%%%%%%%%%%%%%%%%%%%%%%%%%%%%%%%%%%

\end{document}